**Title:**
# Molten Air - A new, highest energy class of rechargeable batteries
**Authors:** S. Licht[1]*

**Affiliation:**

[1]Department of Chemistry, George Washington University, 725 21st St, Washington, DC, 20052.

*Correspondence to:  slicht@gwu.edu.

**Abstract**: Where are the high energy capacity, cost effective batteries urgently needed for a range of medical, transportation and power generation devices, including in greenhouse gas reduction applications such as overcoming the battery driven "range anxiety" of electric vehicles, and increased capacity energy storage for the electric grid? This study introduces the principles of a new class of batteries, rechargeable molten air batteries, and several battery chemistry examples are demonstrated. The new battery class uses a molten electrolyte, are quasi-reversible (rechargeable), and have amongst the highest intrinsic battery electric energy storage capacities. Three examples of the new batteries are demonstrated. These are the iron, carbon and $VB_2$ molten air batteries with respective intrinsic volumetric energy capacities of 10,000, 19,000 and 27,000 Wh liter$^{-1}$. These compare favorably to the intrinsic capacity of the well known lithium air battery (6,200 Wh liter$^{-1}$) due to the latter's single electron transfer and low density limits.

**One Sentence Summary:** A new rechargeable molten air class of battery is introduced, which uses a molten electrolyte, and has amongst the highest intrinsic electric energy storage capacities.

**Main Text:**

High energy capacity, cost effective batteries are urgently needed for a range of applications, including greenhouse gas reduction applications such as overcoming the battery driven "range anxiety" of electric vehicles, and increased capacity energy storage for the electric grid. In this study, the principles of a new class of batteries, rechargeable molten air batteries are introduced, and several battery chemistry examples are demonstrated. The batteries use a molten electrolyte, are quasi-reversible (rechargeable), and have amongst the highest intrinsic battery electric energy storage capacities. In 2008 we introduced a zirconia stabilized $VB_2$ air battery. This 11e$^-$ (eleven electron) per molecule, room temperature, aqueous electrolyte battery has the highest volumetric energy capacity for a battery, with an intrinsic capacity greater than that of gasoline and an order of magnitude higher than that of conventional lithium ion batteries *(1-3)*. The challenge has been to recharge the battery; that is to electrochemically reinsert 11e- into the battery discharge products. Here, this challenge is resolved through the introduction of a new class of <u>molten air</u> batteries.

Other classes of molten electrolyte batteries had been investigated. A molten sulfur battery has been widely studied, particularly for electric car and grid applications *(4, 5)*. During discharge, the battery uses sulfur and sodium (or potassium) for the respective cathode and anode storage materials, and these high temperature molten components are kept from chemically reacting by a solid electrolyte beta alumina separator. Sulfur cells have moderately high capacity. Both the molten and room temperature class of sulfur cathode batteries *(6)*, are limited by the maximum intrinsic capacity of the 2e- per sulfur (2 Faraday/ 32g sulfur). Another class of molten metal electrolyte batteries utilizes an insoluble, dense, molten cathode during discharge situated below a (less dense) molten metal anode floating on a molten electrolyte. Unlike the molten sulfur battery, this latter class does not require a solid electrolyte separator; but to date has lower capacity. An example of this latter class of batteries is the magnesium-antimony battery with a molten halide electrolyte *(7)*.

In this study, rechargeable molten air batteries are introduced, and several battery chemistry examples are demonstrated. The rechargeable molten cathode instead uses oxygen directly from the air, not stored, to yield high battery capacity. The cathode is compatible with multiple electron anodes made possible by our advances in molten salt chemistry. Three examples of the new batteries are demonstrated. These are the iron, carbon and $VB_2$ molten air batteries with respective intrinsic volumetric energy capacities of 10,000, 19,000 and 27,000 Wh liter$^{-1}$. These capacities compare favorably to the intrinsic capacity of the well known lithium air battery (6,200 Wh liter$^{-1}$) due to the latter's single electron transfer and low density limits. As illustrated in Fig. 1 for the iron molten air example of the battery, during charging, iron oxide is converted to iron metal, and $O_2$ is released to the air. During discharge iron metal is converted back to iron oxide. The electrolyte, $Li_2CO_3$, which melts at 723°C is an effective electrolyte.

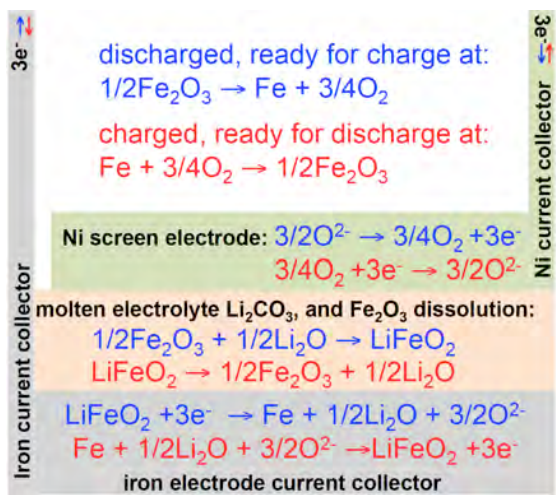

**Fig. 1**. The molten iron air battery. Illustration of the charge/discharge schematic in molten $Li_2CO_3$. Charging process is indicated by red text and arrows, discharge in blue.

We have recently presently a new pathway for the $CO_2$-free synthesis of iron from iron ore based on the electrochemical reduction of iron oxides in molten carbonates *(8-11)*. As shown in the constant current electrolyses in Fig. 2, we have found that iron oxide (as hematite, $Fe_2O_3$, or magnetite, $Fe_3O_4$ readily electrolyzes to iron which forms on an iron or platinum cathode. Post electrolysis, an extracted cathode is shown (left photo) with an overlay of electrolyte; the middle photo shows the cathode after the deposit. The deposit is easily peeled from the cooled cathode, and after peeling, the post electrolysis iron foil cathode is shown (middle foil) ready for a repeat electrolysis. A cross section of the cathode deposit is shown in the right photo, and the iron deposit is evident. The coulombic efficiency of the deposited iron, as well as the composition of iron in the layers adjacent to the cathode are delineated in the supplemental information.

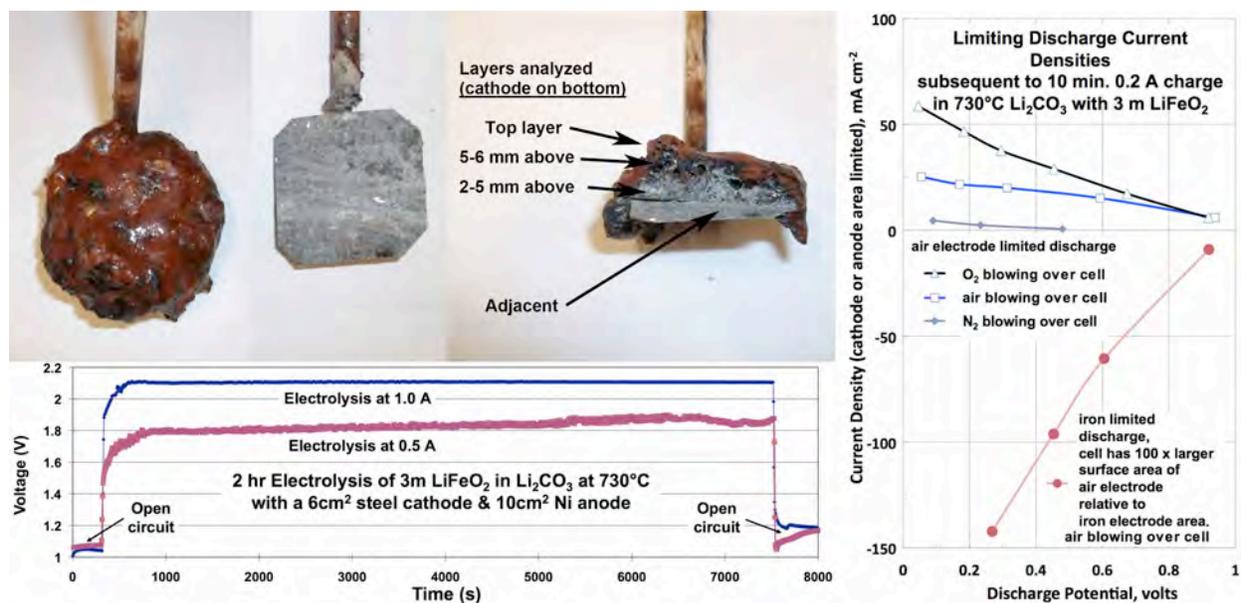

**Fig. 2**. The molten iron air battery. Charging (in this case from dissolved $LiFeO_2$, dissolving 1:1 $Li_2O$ to $Fe_2O_3$) forms a thick iron layer on the cathode as described in the text. Rights side: Discharge polarization (following electrochemical charge to form iron) of the air and iron electrodes in 730°C molten lithium carbonate with $LiFeO_2$.

As evident in the left side of Fig. 2, a steel foil cathode (and a oxygen evolving nickel foil anode) is effective iron for iron oxide splitting, and we have previously shown the wide temperature, concentration and carbonate composition domain in which nickel and iridium are effective air (anode) electrodes and steel, iron and platinum are effective iron (cathode) electrodes for this charging process. These anodes and cathodes can sustain high (mA to A per $cm^{-2}$) current densities at low overpotential in molten lithium carbonate for the oxidation of oxides to oxygen (for either iron oxide (*8-10*) or carbon dioxide splitting (*12,13*)). The same electrolysis shown in the lower portion of Fig. 2 can also represent charging of an iron molten air battery at high current density and low potential. Interestingly, as seen in the right side of Fig. 2, the reverse case, the reduction of oxygen to oxides, is also effective. Ni – air, and steel – iron, electrodes are used, but iridium as the air, and platinum as the iron electrode are also effective and exhibit similar polarization currents. As expected, the oxygen reduction electrode is rate limiting. That is, the oxygen reduction reaction incurs with a several fold higher overpotential (lower current density at equivalent electrolysis potentials) than the cathode iron oxidation discharge reaction. In the absence of oxygen (under nitrogen in the figure), it is seen that the current density diminishes to insignificance, while pure $O_2$, rather than air, increases the current density establishing the basis for electrochemical discharge of the cell.

In 1901, Edison developed rechargeable batteries based on the discharge of an iron anode in (aqueous electrolyte room temperature) to iron oxide (*14*). Retention of an even a small fraction of the intrinsic storage <u>anodic</u> capacity of these batteries has been a challenge, and room temperature iron batteries continue to be explored today (*15*). The 3e⁻ <u>cathodic</u> capacity of iron oxides has also been explored (*16-18*). In 2010 we introduced the molten carbonate electrolytic conversion of iron oxide to iron as a $CO_2$-free alternative to the conventional greenhouse gas intensive industrial production of iron metal. The unexpected high solubility of iron oxide in lithiated molten carbonate electrolytes was demonstrated to lead to the facile splitting of iron oxide to iron metal with the concurrent release of oxygen (*8*). Here, we consider this unusual electrolytic splitting as a battery "charging". We couple this with the known primary discharge of the air cathode as used in the widely studied molten carbonate fuel cell, including those using coal as a fuel (*19,20*), to explore the first example of a molten air rechargeable battery. In lieu of iron, we also explore the alternative use of carbon and $VB_2$ as high capacity discharge anodes for these rechargeable cells. Electrochemical storage in these iron, carbon or $VB_2$ molten air batteries is in accord with:

Iron molten air battery, 3e- discharge/charge: $Fe + 3/4O_2 \rightleftharpoons 1/2Fe_2O_3$ (1)

Carbon molten air battery, 4e- discharge/charge: $C + O_2 \rightleftharpoons CO_2$ (2)

$VB_2$ molten air battery, 11e- discharge/charge: $VB_2 + 11/4O_2 \rightleftharpoons B_2O_3 + V_2O_5$ (3)

The right column of Table 1 compares the intrinsic capacity of these batteries, which is *one to two orders of magnitude greater* than that of the volumetric energy capacity of conventional Li ion batteries. Lithium (metal) air also has a lower volumetric energy capacity (6,200 Wh liter$^{-1}$). Note, that while Li's gravimetric charge capacity (3860 Ah/kg) is similar to that of $VB_2$, it has lower volumetric capacity due to a low density (0.534 kg liter$^{-1}$), and single, rather then multiple, electron charge transfer.

**Table 1**. The intrinsic energy storage capacity of various molten air rechargeable batteries. Note, the cell potential at unit activity, E°, is temperature dependent. For example, while constant over a wide temperature range at 1.0 V for the carbon anode, E° decreases for the iron anode from 1.2 to 0.9 V with temperature increase from 25°C to 850°C (*12*). The volumetric energy capacity, $E_{vol}$, is calculated from the number of electrons stored, n, the density d, the Faraday constant, F = 26.80 Ah mol$^{-1}$, the formula weight, FW, and E° in accord Eqs. 1, 2 or 3 as $E_{vol} = ndE°F / FW$.

| Anode | Formula Weight kg mol$^{-1}$ | Electrons Stored | Charge Capacity Ah/kg | Density kg liter$^{-1}$ | E° versus $O_2$ | Energy Capacity (gravimetric) Wh kg$^{-1}$ | Energy Capacity (volumetric) Wh liter$^{-1}$ |
|---|---|---|---|---|---|---|---|
| Iron | 0.05585 | 3e- | 1,440 | 7.2 | 1.0 | 1,400 | 10,000 |
| Carbon | 0.01201 | 4e- | 8,930 | 2.1 | 1.0 | 8,900 | 19,000 |
| $VB_2$ | 0.07256 | 11e- | 4,060 | 5.1 | 1.3 | 5,300 | 27,000 |

**Iron molten air battery**. The top of Fig. 3 shows various carbonate electrolytes after pressed iron oxide on a cathode is electrolytically reduced, and shows that the air anode is unaffected by the electrolysis. The lower row of photos shows the $Fe_2O_3$ before and after electrolytic reduction and conversion to iron metal. The iron formation in the photograph of Fig. 3 represents charging of the iron molten air (iron/molten electrolyte/air) battery.

The middle and lower portions of Fig. 3 present charge/discharge cycling of an upper, planar 1.0 cm$^2$ nickel gauze (100 mesh) electrode exposed to the air, and separated by 2 g molten lithium carbonate from a lower, 1.0 cm$^2$ planar Pt electrode. Note that lithium oxide, which we have previously shown solubilizes iron oxide, is not added to the molten $Li_2CO_3$ electrolyte. This minimizes the concentration of soluble iron oxide, which can alloy with nickel and diminish cell potential (*10*). Sitting on the platinum is thin, pressed (at 4000 psi), sintered (at 750 °C for 1 hour) $Fe_2O_3$. Platinum was used (figure bottom), rather than steel foil (figure middle), steel screen or coiled iron wire (which are each effective as current collectors), to establish in this experiment that iron from $Fe_2O_3$, rather iron from the current collector, is deposited during charge and oxidized during discharge. We have previously quantified that nickel is highly stable as an air anode in a variety of lithium oxide electrolytes and less so in sodium/potassium carbonate electrolytes. Iridium was used (figure bottom) here, rather than nickel (figure middle), to establish that both are effective as the air electrode for both charge and discharge. The middle portion of the figure presents the voltage of consecutive charge/discharge cycles of the battery when charged at a constant 20 mA and discharged over a constant, 100 ohm load. Alternatively, lithium-free, Ba/Ca/Na/K carbonate electrolytes, or lithium-mix carbonate eutectics, can also be effective as electrolytes for the iron/molten air battery.

The intrinsic capacity of the iron in this example of the molten air battery is high, and higher charge rates (200, rather 20 mA, with greater charging overpotential) are used to expedite probing extended cycling behavior of the cell. As seen in the lower portion of the figure in the first discharge cycle, the battery, unless returned to charge mode, continues to discharge to low potentials. With a discharge cutoff of approximately 0.4V, a highly reversible charge discharge cycle is evident. Enhancements of the morphology and electrocatalytic nature of the air electrode should further improve energy efficiency of the cell.

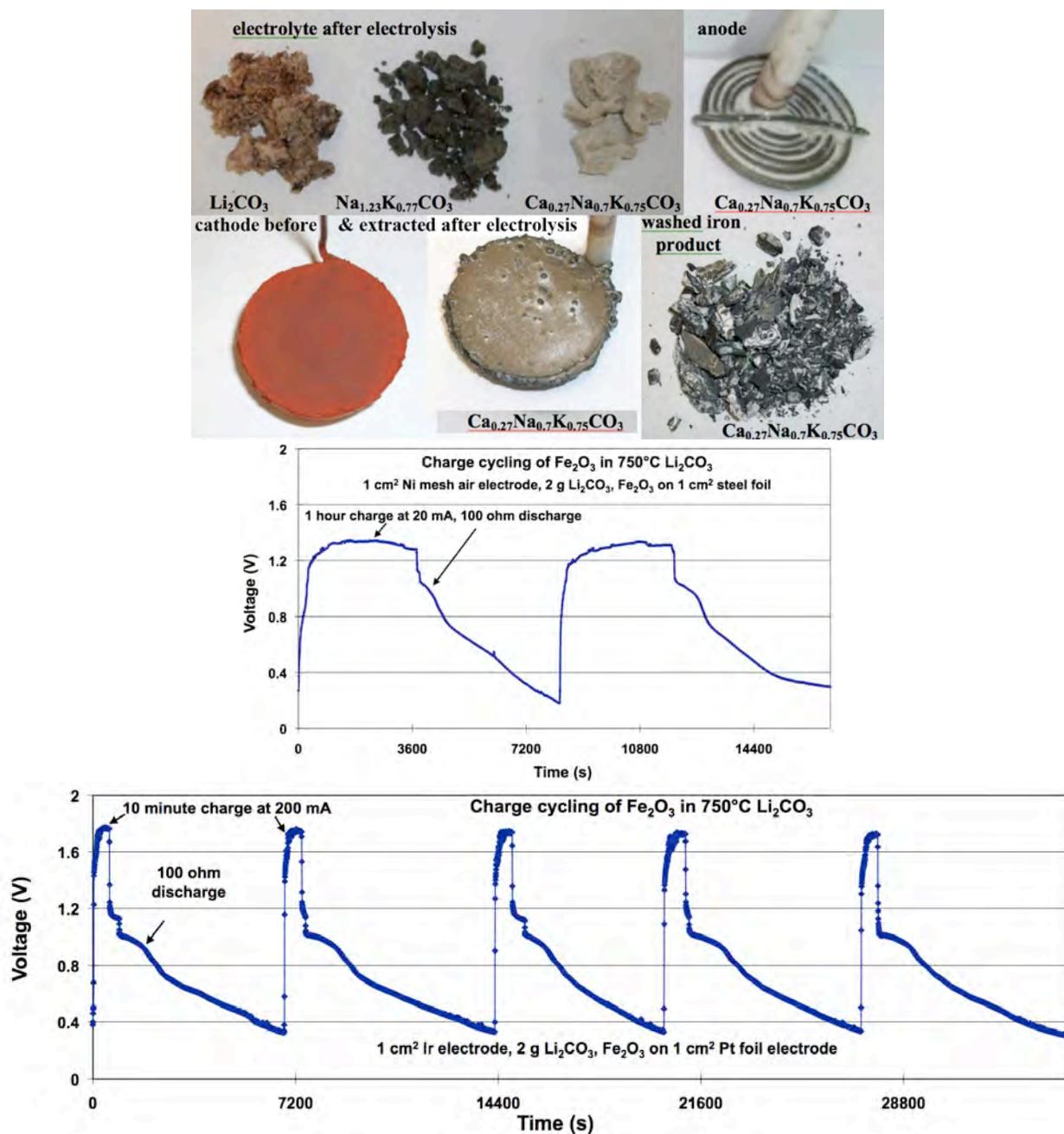

**Fig. 3**. The iron molten air battery. Top: Some carbonates are more effective in promoting an even iron deposition during battery charge. The nickel air anode is highly stable (shown extracted subsequent to electrolysis, with a small coating of solidified electrolyte). Lower row photos: sintered, pressed $Fe_2O_3$ on an iron current collector, prior to electrolysis (left) extracted subsequent to electrolysis (middle) and this product after washing with water (right). Subsequent experiments used a thin cell configuration: nickel screen (100 mesh) as the air electrode and 750°C molten $Li_2CO_3$ electrolyte, and sintered $Fe_2O_3$ on flat platinum or steel foils as the iron electrode, in a 1 cm diameter alumina crucible with vertical walls. Ir or Ni and Pt or steel current collectors exhibit similar results. Charge/discharge cycling at 20 mA (middle) or 200 mA (bottom) constant current followed by discharge at a constant load of 100 ohm.

**Carbonate electrolyte stability**. We have recently shown that under many conditions, molten lithium carbonate readily absorbs atmospheric carbon dioxide, providing a facile route to decrease $CO_2$ in the atmosphere. As summarized in the $CO_2$/carbonate equilibrium diagram in Fig. 4, we now calculate from the known variation of standard enthalpies and entropies with temperature (*21-23*) that barium carbonate has an even larger affinity for $CO_2$ capture than lithium carbonate. The portion above, or below, the equilibrium curves in the figure summarize the partial pressure of $CO_2$ and dissolved oxide concentration relative to carbonate in which $CO_2$ absorption occurs (above the line) or carbonate decomposition occurs (below the line). This equilibrium can be controlled through cation and temperature choice, and the concentration of oxide in the melt. Molten carbonate can gain or lose mass through $CO_2$ absorption or emission, and the concentration of carbon dioxide maintained above the melt is in accord with the equilibrium:

$$MCO_3 \rightleftharpoons MO + CO_2; \quad M= Li_2, Na_2, K_2, Ba, \text{etc} \tag{4}$$

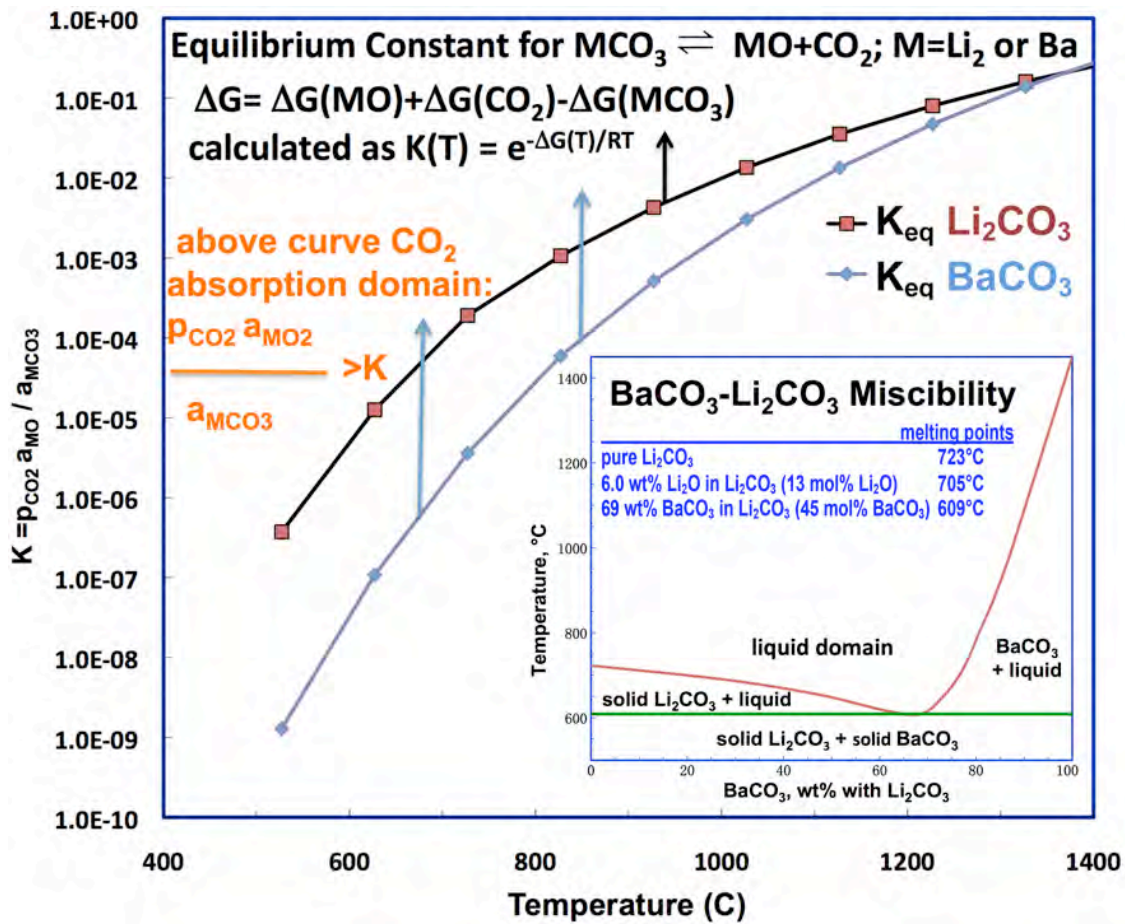

**Fig. 4**. Comparison of the equilibrium constant for $Li_2CO_3$ and $BaCO_3$, decomposition or absorption of $CO_2$ as calculated from the thermochemistry of the carbonate, carbon dioxide and oxide components. Inset: Phase diagram of the $BaCO_3$ - $Li_2CO_3$ system (100% or 0% respectively indicate pure barium or lithium carbonate); modified from reference (*24*). The 705°C $Li_2O/Li_2CO_3$ mix melting point is noted for comparison, but not shown on the curve.

Barium carbonate is solid at lithium carbonate's melting point 723°C. However as summarized in the Fig. 4 inset, barium carbonate readily dissolves in lithium carbonate, and a eutectic comprised of 69% by mass barium to lithium carbonate melts at 609°C. As we have previously observed with electrolyses of lithium carbonate electrolytes (*13*), and as seen in the photo inset of Fig. 5, electrolysis of the barium/lithium carbonate mix at 750° forms a voluptuous layer of carbon on the (steel) cathode side facing the air electrode. When extracted, cooled washed and weighed the carbon mass approaches the theoretical mass calculated by the 100% coulombic efficiency of the four electron reduction of carbonate. The photo is the cathode extracted after 4 hours of 1A electrolysis in the barium mix electrolyte denoted as the pink electrolysis potential versus time plot in the top left portion of Fig. 5. Note at temperatures above 800°C, an increasing portion of carbon monoxide, rather than carbon, would be formed at the cathode.

**Carbon molten air battery**. Carbon formation, as noted above during molten carbonate electrolyses, here represents charging of the carbon/molten air battery. Molten carbonate cells have been widely probed as robust fuel cells (the opposite mode is our electrolysis charging mode), and we had demonstrated that the reverse of this process provides new opportunities for carbon capture. The reversible nature of these two processes provides new opportunities for high capacity battery storage. The top left chart of Fig. 5 summarizes the high rates of charging, at low potential, sustained by the carbon molten air battery using a variety of carbonate electrolyte and cell configurations. Barium carbonate is more facile in $CO_2$ uptake for carbon capture. However, pure lithium carbonate, which also forms a robust carbon overlayer, is more conductive for batteries, and for this carbon molten air battery demonstration we continue with the lithium carbonate electrolyte. To the right of this charge is presented a comparison of the polarization during discharge in a variety of electrolyte and discharge conditions.

The middle and lower portions of Fig. 5 presents charge cycling between an upper, planar 1.0 cm$^2$ nickel gauze (100 mesh) electrode exposed to the air, and separated by molten lithium carbonate from a lower, 1.0 cm$^2$ planar Pt electrode. As with the precious molten air battery example, steel is as effective as platinum as the lower electrode current collector, however platinum is used to demonstrate the current collector metal is not involved in oxidation or reduction, but only as a conduit for charge transfer. A thin cell is used (2 g of 750°C $Li_2CO_3$ horizontally sandwiched between the electrodes) due the high intrinsic capacity of the lithium carbonate in the battery. The middle portion of the figure presents the voltage of consecutive charge/discharge cycles of the battery when charged at 20 mA and discharged over a constant, 100 ohm load. In the lower portion of the figure, high, constant (200 mA or higher current) rate of charge are used to expedite study, and discharge is limited by the rate of (constant resistive load) discharge. The bulk of the discharge occurs as (air reduction and carbon oxidation) at ~0.5V, as seen during the long discharge cycle.

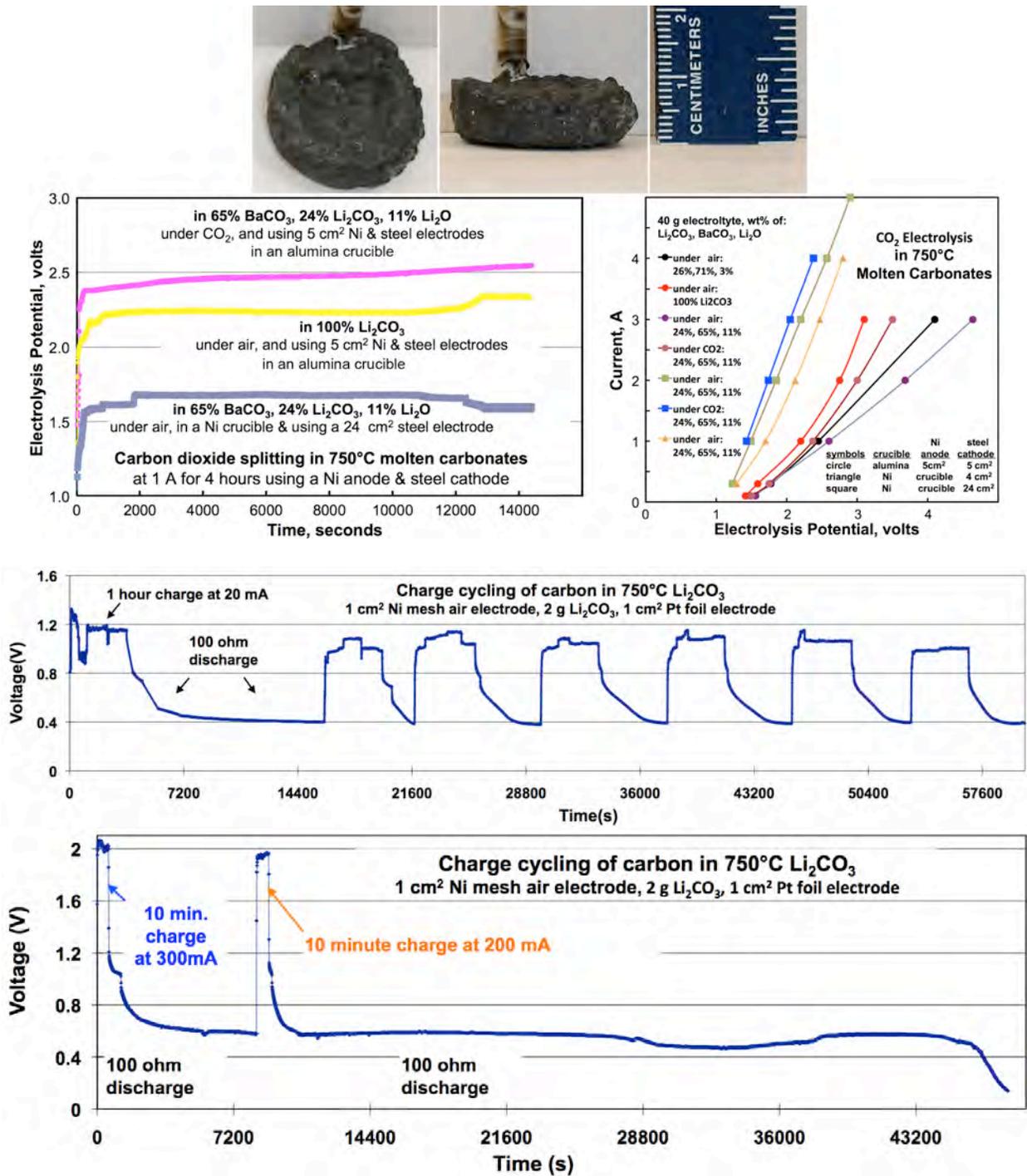

**Fig. 5**. The carbon molten air battery. Photo: The thick carbon layer formed on a 5 cm² steel cathode by a 4 hour, 1 A charge in molten carbonate. Top graphs: Electrolysis potential and charging polarization in various carbonates. Lower graphs: Thin cell charge/discharge cycling at 20 mA (middle) or higher (bottom) constant currents, followed by a 100 load ohm discharge. Thin cell configuration: air electrode-Ni screen, electrolyte-2g 750°C molten $Li_2CO_3$, carbon deposition electrode- Pt, in a 1 cm diameter alumina crucible with vertical walls. Ir or Ni and Pt or steel current collectors exhibit similar results.

**VB$_2$ molten air battery**. While there are few studies regarding <u>charging</u> of carbonates, there is a wide body of knowledge regarding the electrochemistry of <u>discharge</u> in molten carbonates (the molten carbonate fuel cell literature). In comparison, the electrochemical foundation of understanding of the VB$_2$ molten air system is even smaller, and there is little in the way of prior studies regarding the electrochemistry of the Eq. 3 discharge products - molten vanadates or molten borates. Here, we provide a path towards recharge of the VB$_2$ molten air battery. The studies are less advanced due to the scarcity of prior fundamental electrochemical knowledge of the system.

The VB$_2$ discharge products B$_2$O$_3$ (mp 450 °C, white, melts clear) and V$_2$O$_5$ (mp 690 °C, yellow/brown) have a low melting point compared to VB$_2$ (mp 2450 °C, black). The molten B$_2$O$_3$ and V$_2$O$_5$ salts are miscible, whereas VB$_2$ does not appear to be miscible , and due to is higher density VB$_2$ descends in the molten mixture. However, this molten phase is an insulator, without significant ionic dissociation. The melt cannot be electrochemically charged. Dissolution of an oxide, such as Li$_2$O or, in the melt provides adequate ionic conductivity, and charging of the molten electrolyte occurs. The high temperature (molten) phase of the Li$_2$O-B$_2$O$_3$-V$_2$O$_5$ system has not been previously explored, but ionic conductivity range of this system in the solid phase has been established at temperatures up to 250°C (*25*). The higher temperature binary system B$_2$O$_3$ (mp 450 °C) and Li$_2$O (mp 1438 °C, white, dissolves clear) presents a complex phase diagram with an extensive homogenous liquid phase above 767 °C (*26*). Domains of this diagram that we have probed are indicated by colored lines in the top of Fig. 6.

For this study we have explored both single cell and separated half cell (air electrode in a crucible with a porous bottom sitting in the electrolyte of a larger crucible with the VB$_2$ electrode) configurations, and have found no advantage of the more complex separated cell configuration. The alumina (Al$_2$O$_3$) crucible is susceptible to corrosion in the high Li$_2$O concentration electrolytes (to LiAlO$_3$) particularly at temperatures above 800°C. Lithiated alumina ceramic crucibles would be preferred, but are unavailable. Charging of the electrolyte of the B$_2$O$_3$, V$_2$O$_5$, Li$_2$O mix electrolyte is accomplished at a coiled nickel wire anode above a coiled steel wire cathode in an alumina crucible, and yields a thick black deposit on the cathode. A 2A charge for 6 hours results in a mass loss of 1.5 g, only ~40% of the expected mass loss due to oxygen evolution via equation 3. Alternative means to improve this low charging coulombic efficiency will be explored in future experiments. We hypothesize that this low charging efficiency is due to reoxidation of the product by oxygen dissolved within the electrolyte (hence, the brief foray into the separated half cell configuration). Alternatively, equivalent mixtures of LiVO$_3$ and LiBO$_2$ yield similar results. Following extraction, cooling, and acid wash to remove soluble components, FTIR of the product exhibits absorption peaks at 1640 and 3470 cm$^{-1}$ that overlap with that of commercial (NOAH high purity chemicals) VB$_2$. Fig. 6 presents the liquid domain and complexity of the binary Li$_2$O : B$_2$O$_3$ (wt%) system, a photo of the VB$_2$ on steel wire cathode, extracted following charging, discharging of thin VB$_2$ molten air batteries with various electrolytes including an experimental CaO electrolyte, and cycling of the VB$_2$ molten air battery under heavy (1 ohm) load.

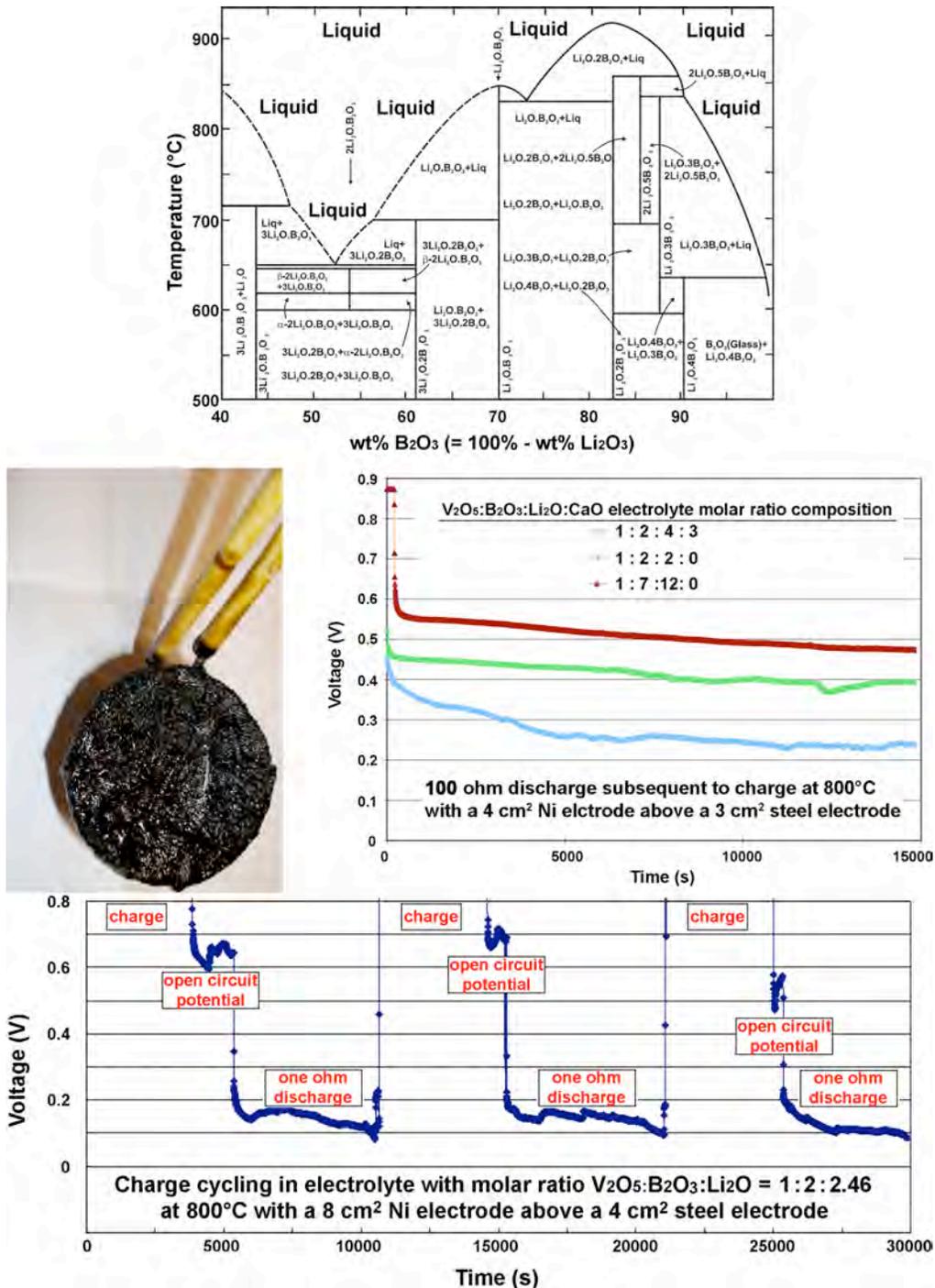

**Fig. 6**. The VB$_2$ molten air battery. Top: the liquid component of the binary Li$_2$O : B$_2$O$_3$ (wt%) system (modified from reference 26). Photo: VB$_2$ on steel wire cathode, extracted following an 8.5 hour charge at 0.2 A in a 767°C electrolyte with a molar ratio of Li$_2$O : V$_2$O$_5$ : B$_2$O$_3$ = 1 : 2 : 0.67 in the separated half cell configuration described in the text. Middle: thin cell cell configuration (not separated), 100 ohm discharge potential subsequent to a 0.2A charging for 10 minute, in 2 g of various 800°C vanadate borate electrolytes sandwiched between a coiled disc nickel wire electrode above and a coiled disc steel wire electrode below. Bottom: Rapid (1 ohm) cycled discharge at 800°C in, V$_2$O$_5$, B$_2$O$_3$, Li$_2$O electrolyte.

A proof of principle of the rechargeable $VB_2$ molten air system has been presented, starting with a cell in the discharged (molten borate and vanadate) condition. Impediments to efficiency are recombination of the solid product formed at the cathode with solution phase oxygen, and the poor conductivity of the cathode product inhibiting discharge.

**Conclusions**:

The foundations and experimental demonstration of a new class of molten air batteries is extablished. The iron molten and carbon molten air batteries exhibit high rate charging capability and quasi reversibility (rechargeability). An extensive range of experimental parameters can be investigated to start the optimization process of these batteries. For example, while these experiments have been conducted in the 700°C temperature range, the molten carbonate electrolyte has a wide range of electrolyte opportunities, and mixed alkali carbonate eutectics have a minimum melting point below 400°C. A range of cell configurations with lower polarization (with similar discharge potentials, but supporting significantly higher current density) will be reported on in a future study.

**References:**

1. S. Licht, H. Wu, X. Yu, Y. Wang, Renewable Highest Capacity $VB_2$/Air Energy Storage. *Chem. Comm.* **2008**, 3257-3259 (2008).

2. S. Licht, S. Ghosh, B. Wang, D. Jiang, J. Asercion, H. Bergmann, Nanoparticle facilitated charge transfer and voltage of a high capacity $VB_2$ anode. *Electrochem. Solid State Lett.* **14**, A83-A85 (2011).

3. S. Licht, C. Hettige, J. Lau, U. Cubeta, H. Wu, J. Stuart, B. Wang, Nano-$VB_2$ synthesis from elemental vanadium and boron: nano- $VB_2$ anode/air batteries. *Electrochem. Solid State Lett.* **15**, A12-A14 (2012).

4. B. Dunn, H. Kamath, J. M. Tarascon, Electrical Energy Storage for the Grid: A Battery of Choices. *Science* **334**, 9280935 (2011).

5. J. B. Goodenough, Rechargeable batteries: challenges old and new, *J. Solid State Electrochem.* **16**, 2019-2029 (2012).

6. D. Peramunage, S. Licht, A Novel Solid Sulfur Cathode for Aqueous Batteries. *Science*, **261**, 1029-1032 (1993).

7. D. J. Bradwell, H. Kim, A. H. C. Sirk, D. R. Sadoway, Magnesium-Antimony Liquid Metal Battery for Stationary Energy Storage. *J. Amer. Chem. Soc.* **134**, 1895-1897 (2012).

8. S. Licht, B. Wang, High Solubility Pathway to the Carbon Dioxide Free Production of Iron. *Chem. Comm.* **46,** 7004-6 (2010).

9. S. Licht, H. Wu, Z. Zhang and H. Ayub, Chemical Mechanism of the High Solubility Pathway for the Carbon Dioxide Free Production of Iron. *Chem. Comm.* **47,** 3081-3083.

10. S. Licht, H. Wu, STEP iron, a chemistry of iron formation without $CO_2$ emission: Molten carbonate solubility and electrochemistry of iron ore impurities, *J. Phys. Chem., C,* **115**, 25138-25147 (2011).

11. S. Licht, Efficient Solar-Driven Synthesis, Carbon Capture, and Desalinization, STEP: Solar Thermal Electrochemical Production of Fuels, Metals, Bleach. *Advanced Materials* **47**, 5592-5612 (2011).

12. S. Licht, STEP (solar thermal electrochemical photo) generation of energetic molecules: A solar chemical process to end anthropogenic global warming. *J. Phys. Chem. C,* **113**, 16283-16292 (2009).

13. S. Licht, B. Wang, S. Ghosh, H. Ayub, D. Jiang, J. Ganley, A New Solar Carbon Capture Process: Solar Thermal Electrochemical Photo (STEP) Carbon Capture. *J. Phys. Chem Lett.* **1**, 2363-2368 (2010).

14. T. A. Edison, Reversible Galvanic Battery. *US Patent* 678722 (1901); *ibid*, 692507 (1902).

15. A. K. Manohar, S. Malkhandi, B. Yang, C. Yang, G. K. S. Prakash, S. R. Narayanan, A High-Performance Rechargeable Iron Electrode for Large-Scale Battery Based Energy Storage. *J. Electrochem. Soc.* **159**, A1209-A1214 (2012).

16. S. Licht, B. Wang, and S. Ghosh, Energetic Iron(VI) Chemistry: The Super-Iron Battery. *Science* **285**, 1039-1042 (1999).

17. S. Licht, A High Capacity Li-ion Cathode: The Fe(III/VI) Super-iron Cathode. *Energies* **3**, 960-972 (2010).

18. M. Farmand, D. Jiang, B. Wang S. Ghosh, D. Ramaker, S. Licht, Super-iron nanoparticles with facile cathodic charge transfer. *Electrochem. Comm.* **13**, 909-912 (2011).

19. A. C. Rady, S. Giddey, S. P. S. Badwal, B. P. Ladewig, S. Bhattacharya, Review of Fuels for Direct Carbon Fuel Cells. *Energy & Fuels* **26**, 1472-1488 (2012).

20. A. Kulami, S. Giddey, Materials issues and recent developments in molten carbonate fuel cells. *J. Solid State Electrochem.* **16**, 3123-3146 (2012).

21. M. W. Chase, *J. Phys. Chem. Ref. Data* **9**, 1 (1998).

22. U.S. NIST ChemWeb online thermochemical data available at:
   http://webbook.nist.gov/chemistry/form-ser.html

23. Glenn Research Center NASA, *ThermoBuild access to NASA Glenn thermodynamic CEA database*, 2006; data available at:
   http://www.grc.nasa.gov/WWW/CEAWeb/ceaThermoBuild.htm

24. P. Pasierb, R. Gajerski, M. Rokita, M. Rekas, Studies on the binary system $Li_2CO_3$-$BaCO_3$. *Physica B* **304**, 463-476 (2001).

25. Y. Lee, J. Lee, S. Hong, Y. Park, Li-ion conductivity in $Li_2O$–$B_2O_3$–$V_2O_5$ glass system. *Solid State Ionics* **175** 687–690 (2004).

26. E. B. Ferreira, E. Zanotto, S. Feller, G. Lodden, J. Banerjee, T. Edwards, and M. Affatigat, Critical Analysis of Glass Stability Parameters and Application to Lithium Borate Glasses. J. Am. Ceram. Soc., 94, 3833–3841 (2011).

**Acknowledgments:**

The author is grateful to the support of the United States National Science Foundation for support of this work.

**Supplementary Materials:**

Materials and Methods (at end)

Figures S1-S24

Tables S1-S21

References R1-R4

**Supplementary Materials: for**
**Molten Air - A new, highest energy class of rechargeable batteries** - S. Licht
Department of Chemistry, George Washington University, 725 21st St, Washington, DC, 20052.

Materials and Methods (at end)

Figures S1-S24, Tables S1-S21, References R1-R4

**Supplementary Materials:**

We had recently discovered that the lithiation of iron oxides facilitates their concentrated dissolution in molten carbonates. Both hematite, $Fe_2O_3$, and magnetite, $Fe_3O_4$, are highly soluble in molten lithiated carbonates (*8-11*). We have observed a high solubility for both lower temperature eutectic carbonate melts ($Li_{0.87}Na_{0.63}K_{0.50}CO_3$), and for pure $Li_2CO_3$, (pure $Li_2CO_3$ melts at 723°C). In $Li_2CO_3$, ferric, [Fe(III)], solubility increases from 7 to 12 molal (molal = m ≡ moles per / kg $Li_2CO_3$) with temperature increase from 750 to 900°C, while in the eutectic the solubility increases from 1 to 4 m Fe(III) as temperature increases from 550°C to 900°C.

Lithium oxide is not consumed in the iron making process. For $Fe_2O_3$ in molten carbonate, the STEP iron production mechanism is given by (*9*):

**I** dissolution in molten carbonate 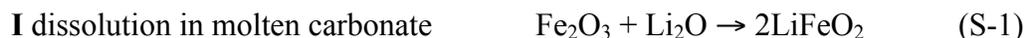 $Fe_2O_3 + Li_2O \rightarrow 2LiFeO_2$ (S-1)

**II** electrolysis, $Li_2O$ regeneration: 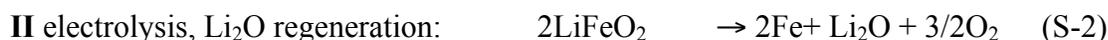 $2LiFeO_2 \rightarrow 2Fe + Li_2O + 3/2O_2$ (S-2)

**III** net reaction: 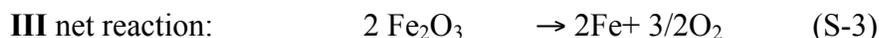 $2Fe_2O_3 \rightarrow 2Fe + 3/2O_2$ (S-3)

*Speciation profile of iron deposited during charge.*

Fig. S-1 profiles the measured concentrations of the reduced iron species subsequent to electrolysis. The width reflects the relative total fraction of iron metal found in each layer. The vertical subdivisions of each bar reflect the relative iron oxide concentration that has been reduced to $Fe^{2+}$ or to iron metal. Product closer to the cathode is on the left side, and product further from the cathode is on the right side of the scheme.

The Fig. S-1 compositional analysis provides an understanding of each of the layers evident in the photograph of the cathode product. Consistent with the grey-metal color, the dominant iron species in the layer adjacent to the cathode is zero-valent iron. The additional black evident in the next layers is consistent with the color of partially reduced iron oxides, such as ferrous oxide or $Fe_3O_4$. The thin, top layer on the product was still molten while the cathode was extracted. The red/brown color of this layer is similar to that of the remaining bulk electrolyte, but still contains a significant (zero valent) iron metal content. Finally although not included in this mapping and furthest from the cathode, the post electrolysis bulk electrolyte also contains zero valent iron (although at a much lower level, ~4% of that found at the cathode surface), as well as partially reduced (ferrous) iron, and also ferric ion remaining from the initial iron oxide dissolution.

The product speciation profile in Fig. S-1 presents the substantial, but incomplete reduction of iron. One objective of this supplemental study is the systematic optimization of the STEP Iron to approach 100% coulombic efficiency of the conversion of ferric ion, $Fe^{3+}$, to iron metal, $Fe°$, and a second objective to accomplish that high coulombic efficiency at low energy (low applied electrolysis potential and high current density). The electrolysis of Fig. 2 (main paper), is conducted with parallel, horizontal anode and cathode. The nickel oxygen evolving anode is situated near the surface of electrolyte, 1 cm above the cathode. The 10 $cm^2$ nickel electrode is oversized compared to the 6.2 $cm^2$ cathode.

The electrolyte composition and temperature, and the electrolysis configuration, duration and current density have a substantial impact on STEP Iron coulombic efficiency. For example, the electrolysis of Fig. 2 is conducted at 730°C, near the electrolyte melting point, and a rise in temperature to 850°C decreases the requisite electrolysis potential but generates a lower fraction of iron metal. This supplementary material in the next section begins with an electrolysis temperature of 800°C, and systematically conducts modification of the physical chemical components of the system to understand conditions which increase coulombic efficiency and minimize the requisite electrolysis potential.

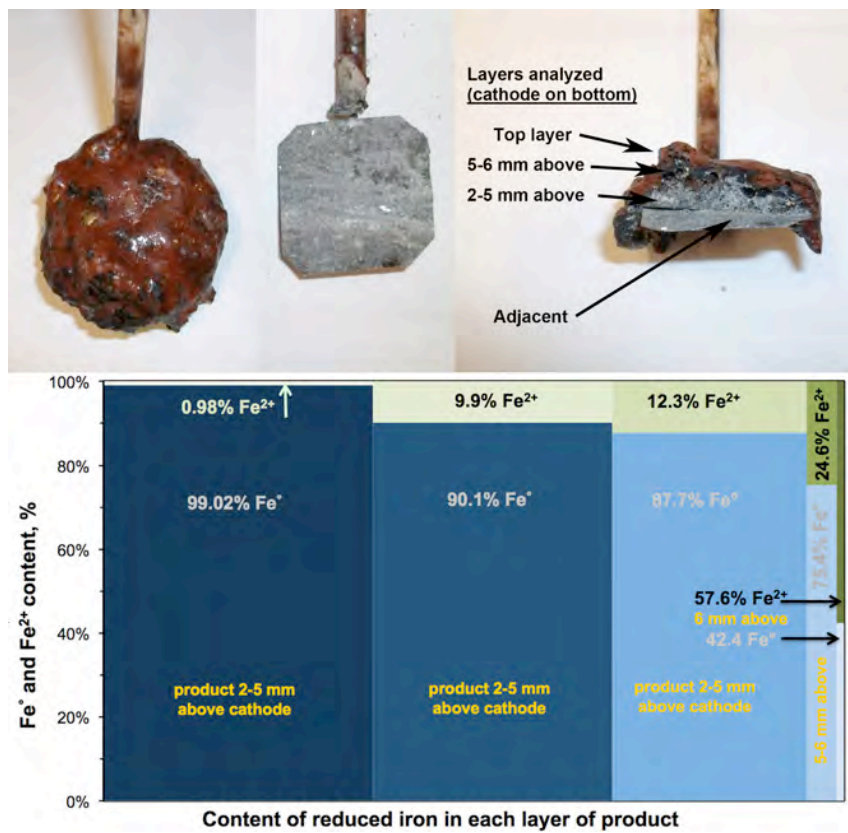

**Fig. S-1**. Bottom: Cross sectional profile of the reduced iron species analyzed from the layers (right photo) subsequent to cathode extraction. The layers analyzed are shown in the righthand photo. Deposition conditions are given in Fig. 2 (main paper). Post electrolysis, the extracted cathode is shown (left photo) with an overlay of electrolyte; the middle photo shows the cathode after the deposit. The deposit is easily peeled from the cooled cathode, and after peeling, the post electrolysis iron foil cathode is shown (middle foil) ready for a repeat electrolysis.

Page 17

Following extensive modification and systematic optimization as delineated in the next section, optimized, high efficiency, low energy electrolyses are measured with an outer foil or wire cylinder pure nickel anode and a small or large diameter inner steel foil cathode. The coulombic efficiency compares the applied electrolysis current to the theoretical required for the three electron reduction of $Fe^{3+}$ to iron metal. Efficiency is highest at minimum molten lithium carbonate temperature, intermediate dissolved ferric concentration (3 molal), intermediate cathode and anode current density, anode electron transfer facilitated by situating the anode near the electrolyte//air interface, and or a vertical cathode parallel to a vertical anode. Extended duration electrolyses can be sustained if levels of ferric are not depleted, such as would be achieved via continuous addition of ferric oxide during the electrolysis. The electrolysis configuration is simplified when the electrolysis is conducted in a nickel crucible, which comprises both the anode and the cell body in one piece.

The overpotential is constrained by the anode current density; energy minimization is achieved with a maximum anode surface area. Interelectrode separations must be small enough (< 1 cm) to minimize potential loss at high (> 0.1 A cm$^{-2}$) current density. Consistent with the expected Nernst decrease of redox potential with increasing ferric concentration, higher $LiFeO_2$ (and $LiO_2$) concentration decreases the electrolysis potential

As delineated in the next section, optimized electrolyses occur in a low electrolysis potential of 1.35 V and at coulombic efficiency approaching 100%. The energy required to generate a mole of iron is 1.35V x (3 Faraday / mole Fe) x (26.8 Ah per Faraday) = 109 Wh per mole of iron. This is only approximately 1/3 of the energy required to produce a mole of iron using coke carbon by the conventional carbothermal method (this assumes solar thermal energy is used to maintain the 730°C STEP Iron electrolysis temperature).

Lithium containing carbonate electrolytes promote soluble iron oxide, as mediated by lithium oxide, Eq. S-1. The high solubility leads to low energy iron metal electrodeposition, but the iron product can be spatially diffuse. The carbon footprint of molten carbonate electrolytes with less soluble or insoluble is explored here as green, low cost alternatives to a pure lithium carbonate electrolyte. Iron oxide solubility in a molten carbonate electrolyte decreases when there is less lithium present; the solubility of $Fe^{3+}$ in a $Li_{0.87}Na_{0.63}K_{0.5}CO_3$ eutectic is less than half of its value in $Li_2CO_3$ (*8*).[1] Electrolysis of $Fe^{3+}$ dissolved in this eutectic diminishes the secondary $Fe^{2+}$ layer in the product, compared to that in $Li_2CO_3$, as seen in Fig. S-7.

*Systematic Optimization of electrolytic iron production in molten carbonate.*

A systematic, variation of molten iron electrolysis can yield efficient iron production efficiency at low energy. General conditions of the first series of electrolyses are summarized in Table S-1. The cathode in this first series of experiments is a thin planar 6.25 cm$^2$ steel sheet. This cathode lies under the anode, and we observe that the iron product is deposited on top of the cathode. This cathode surface is the active area exposed to short ion diffusion path between the electrodes.

Table S-2 summarizes the results of 1 hour electrolyses at 1 Amp in an 800°C molten lithium carbonate electrolyte containing 1.5 m $Fe_2O_3$ and 3 m $Li_2O$ (to generate 3 m $LiFeO_2$ in solution). In this series of experiments the anode, a coiled pure nickel (McMaster 200 Ni) wire is situated 3 mm below the surface of the electrolyte to facilitate oxygen evolution in an attempt to minimize oxygen interaction with the cathode product.

**Table S-1.** Constant characteristics in the first series of STEP Iron parametric optimization studies.

| Current of electrolysis | 1.0 amp |
|---|---|
| Time of electrolysis (1h) | 3600 s |
| Theoretical max mass $Fe^0$ from electrolysis: $Fe^{3+} + 3e^- \rightarrow Fe^0$ | 1A×3600÷96485÷3×55.85 = 0.696 g |
| Anode: Ni wire tightly coiled, l = 16 cm, d = 2.0 mm, area: | 10 cm$^2$ |
| Cathode: Steel shim, 2.5cm×2.5cm, area | 6.25 cm$^2$ |

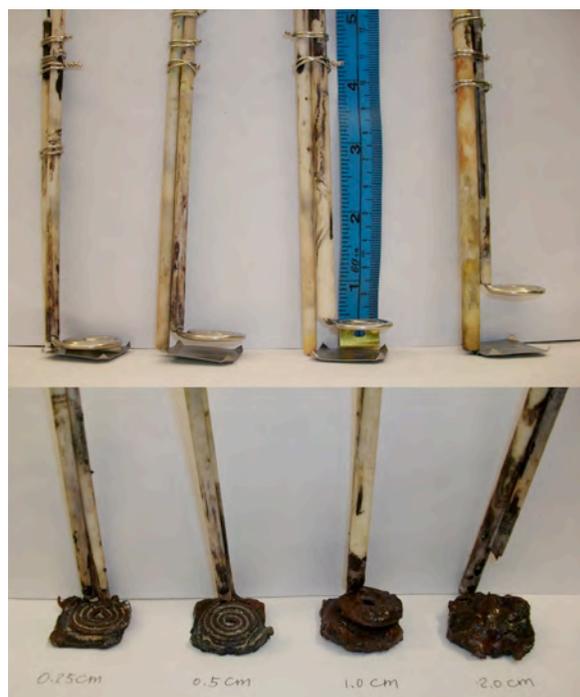

**Fig. S-2**. Anode (upper electrode) and cathode (lower electrode) prior to electrolysis (top) and following electrolysis prior to removal of cathode product (bottom). Insulating alumina ceramic tubes are reused and hence appear charred.

In the first series of experiments, summarized in Table S-2, the anode to interelectrode spacing is varied, and the mass of electrolyte was changed to cover smaller or larger inter-electrode separations. The electrodes before and after electrolysis are photographed in Fig. S-2. The lithium based electrolyte is highly conductive, and even at these relatively high current densities, electrolytic resistance losses are not significantly impacted by the variation of inter-electrode spacing. The electrolysis potentials at 1 A constant current are the same to within 0.1 V, independent of the 0.25 to 2.0 cm electrode separation. Coulombic efficiency, particularly during the 0.25 cm separation electrolysis may have been impacted by shorting as the iron deposit growing from the cathode approached the anode, and that a maximum of over 50% coulombic efficiency is achieved for an intermediate spacing of 1.0 cm.

Table S-3 summarizes STEP Iron electrolyses in which the starting concentration of Li$_2$O is varied. In each case, the initial ferric concentration (as added Fe$_2$O$_3$) is kept constant at 3 molal Fe$^{3+}$ in Li$_2$CO$_3$. As we have previously demonstrated, Fe$_2$O$_3$ is not soluble in carbonate unless Li$_2$O is added, and reacts to form LiFeO$_2$ in the molten solution (*8-10*). The Li$_2$O is not consumed in the electrolysis process. That is consistent with eq. 7, as LiFeO$_2$ is reduced to form iron metal, Li$_2$O is liberated to dissolve the next iteration of added iron ore (Fe$_2$O$_3$). As seen in the photo in Fig. S-3, the iron is deposited directly on top of the cathode, followed by a black layer of partially reduced iron (magnetite), followed by a layer of the electrolyte containing the excess (brown) $Fe_2O_3$ dissolved in the electrolyte. The iron layer is easily separated from the cathode, Fig. S-4, and as seen in Figs. 3 and 4 leaves behind a reusable cathode. As seen in Table S-3, a 1:1 ratio of $Li_2O$ to $Fe_2O_3$ supports the maximum coulombic efficiency, although it is interesting to note in Fig. S-5, that higher concentrations of $Li_2O$ significantly decrease the required electrolysis potential (as seen comparing the 1.5 m and 9.0 m $Li_2O$ electrolysis) consistent with our recent observation (*10*). Also evident in the figure, potential variations are occasionally observed during individual electrolyses, although average electrolysis potential trends are highly consistent.

**Table S-2.** Effect of electrode spacing and electrolyte mass on electrolysis. Electrodes are delineated in Table S-1. The electrolyte mass was changed as indicated below, to maintain electrolyte converage in experiments with smaller or larger inter-electrode separations. Note, coulombic efficiency, particularly during the 0.25 separation electrolysis may have been impacted by shorting as the iron deposit grew from the cathode towards the anode.

| Temperature (°C) | 800 | 800 | 800 | 800 |
|---|---|---|---|---|
| Anode/Cathode separation (cm) | 0.25 | 0.5 | 1.0 | 2.0 |
| Electrolyte Total mass (g): Electrolyte weighed from a mix of 200.0015g $Li_2CO_3$, 47.8888g $Fe_2O_3$ and 17.9286g $Li_2O$, | 8.3193 | 16.6128 | 33.2174 | 66.4510 |
| $Fe^{3+}$ concentration (mol /kg $Li_2CO_3$) as $Fe_2O_3$ | 3.0 m | 3.0 m | 3.0 m | 3.0 m |
| $Li_2O$ concentration (mol /kg $Li_2CO_3$) | 3.0 m | 3.0 m | 3.0 m | 3.0 m |
| Cathode: Steel shim, 2.5cm×2.5cm | 6.25 cm$^2$ | 6.25 cm$^2$ | 6.25 cm$^2$ | 6.25 cm$^2$ |
| $Fe^0$ mass in product (g) | 0.135 | 0.328 | 0.381 | 0.249 |
| Coulombic efficiency (100%x $Fe^0$ mass experiment/theory) | 19.4 | 47.0 | 54.7 | 35.7 |

The next series of experiments utilizes the general conditions described in Table S-2, still at constant (1.0 amps) current, but varies the electrolysis time. As summarized in Table S-4, whereas 1 hour of electrolysis should theoretically yield 0.7 g of iron (assuming 1000% coulombic efficiency of the three electron reduction of dissolved $Fe^{3+}$), 8 hours of electrolysis would be expected to generate 5.6 g of iron metal. As summarized in Table S-4, the experimental coloumbic efficiency during electrolysis is approximately 50 percent and is not substantially affected by the electrolysis time. What is affected, as shown in the photos in Fig. S-6, is the amount of salt that is removed with the product, and is enriched 5-fold in iron metal after the extended electrolysis. The amount of salt accompanying the cathode product is seen to be much lower after 8h electrolysis than after 1 hour electrolysis. In each case, the cathode product is washed prior to the iron(0) content analysis. Even subsequent to extended electrolysis time, each of the cathodes remain intact after the removal of product. As seen in Fig. S-6, the 8 hour electrolysis cathode product does not exhibit the multiple layers evident, and instead has iron metal throughout the product (but the iron remains spatially diffuse intermingled with electrolyte).

**Table S-3.** Effect of Li$_2$O concentration on electrolysis. The anode/cathode separation is 1.0 cm; other electrolysis conditions are detailed in Table S-1. Note, an intermediate Li$_2$O concentration (2 molal) conducted for double the electrolysis time, generated greater iron metal product.

| Temperature (°C) | 800 | 800 | 800 | 800 |
|---|---|---|---|---|
| Time of electrolysis (h) at 1.0 amp | 1h | **2h** | 1h | 1h |
| Theoretical max mass Fe$^0$ from electrolysis: Fe$^{3+}$ + 3e$^-$ → Fe$^0$ | 0.696 g | 1.389 g | 0.696 g | 0.696 g |
| Average Potential of electrolysis (V) | 1.84 | 1.80 | 1.69 | 1.57 |
| Carbonate Electrolyte: Li$_2$CO$_3$ (g) | 25.000 | 25.001 | 24.992 | 21.999 |
| Fe$^{3+}$ concentration (mol /kg Li$_2$CO$_3$) | 3.0 m | 3.0 m | 3.0 m | 3.0 m |
| Fe$_2$O$_3$ mass (g) | 5.9866 | 5.9868 | 5.9843 | 5.9167 |
| Li$_2$O concentration (mol /kg Li$_2$CO$_3$) | 1.5 m | **2.0 m** | **3.0 m** | **9.0 m** |
| Li$_2$O mass (g) | 1.1200 | 1.4941 | 2.241 | 5.9167 |
| Electrolyte Total Weight (g) | 32.1138 | 32.4217 | 33.2174 | 33.1841 |
| Fe$^0$ mass in product (g) | 0.312 | 0.711 | 0.381 | 0.155 |
| Coulombic efficiency (100%x Fe$^0$ mass exper/theory) | 44.8 | 51.7 | 54.7 | 22.3 |

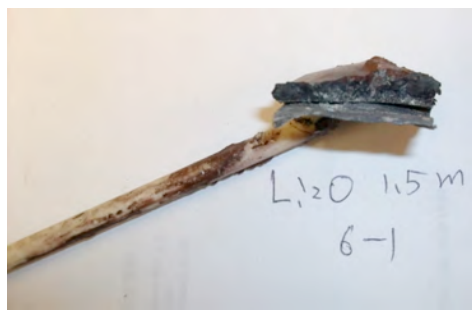

**Fig. S-3.** Side view of cathode with product.

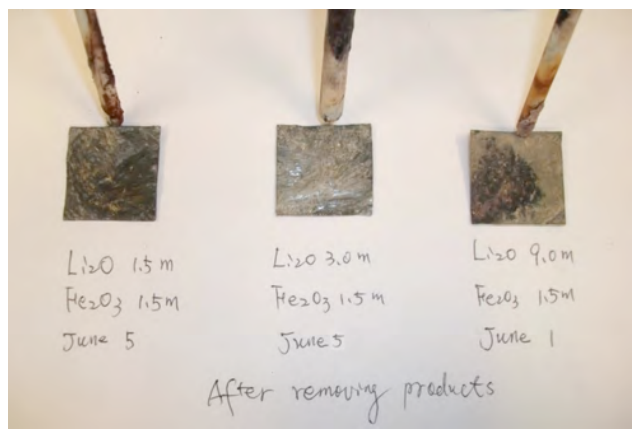

**Fig. S-4.** Cathodes, after removal of product, remain intact.

**Fig. S-5.** Variation of STEP Iron electrolysis potential with different Li$_2$O concentrations.

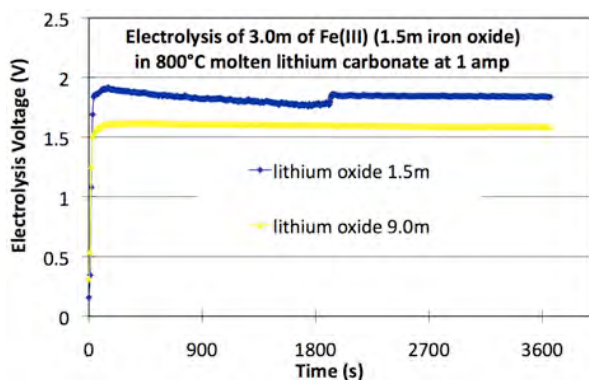

**Table S-4.** Effect of electrolysis time on the electrolytic formation iron. The anode/cathode separation is 1.0 cm; other electrolysis conditions are detailed in Table S-1.

| Temperature (°C) | 800 | 800 | 800 | 800 | 800 |
|---|---|---|---|---|---|
| Time of electrolysis (h) at 1.0 amp | *1h* | *1h 20m* | *2h* | *4h* | *8h* |
| Theoretical max mass Fe$^0$ via: Fe$^{3+}$+3e$^-$ → Fe$^0$ | 0.696 g | 0.926 g | 1.389 g | 2.778 g | 5.557 g |
| Average Potential of electrolysis (V) | 1.69 | 1.75 | 1.57 | 1.7 | 1.65 |
| Fe$^{3+}$ concentration (mol /kg Li$_2$CO$_3$) as Fe$_2$O$_3$ | 3.0 m | 3.0 m | 3.0 m | 3.0 m | 3.0 m |
| Li$_2$O concentration (mol /kg Li$_2$CO$_3$) | 3.0 m | 3.0 m | 3.0 m | 3.0 m | 3.0 m |
| Electrolyte Total Weight (g) | 33.2174 | 33.0001 | 33.2163 | 33.2143 | 33.2177 |
| Fe$^0$ mass in product (g) | 0.381 | 0.529 | 0.732 | 1.493 | 2.553 |
| Coulombic efficiency (100%x Fe$^0$ mass exp/theory) | 54.7 | 57.1 | 52.7 | 53.7 | 45.9 |
| * = anode contact broke just prior to completion | | * | * | | |

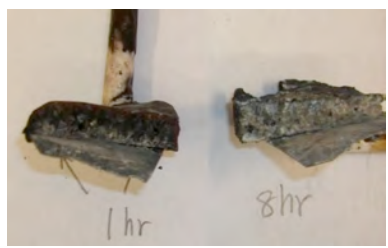

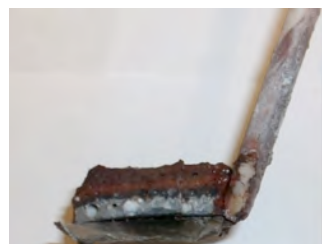

**Fig. S-6**. Product removed and partially peeled from cathode after 1 or 8 hour electrolyses. The amount of salt removed with the product is high after 1h electrolysis (then washed prior to analysis) & lower after the extended (8h) electrolysis. During the 8h electrolysis, a higher fraction of the initial 3 m of Fe$^{3+}$ is converted to Fe' & the product layer sitting on the cathode is enriched in iron metal.

**Fig. S-7**. (right) Side view of cathode with product after 1h 3 m LiFeO$_2$ in Li$_{0.85}$Na$_{0.61}$K$_{0.54}$CO$_3$ 750°C STEP iron electrolysis.

Sodium carbonates (mp 851°C) or potassium carbonate (mp 891°C) both have higher melting points than lithium carbonate (mp 723°C). However, a eutectic mix of the three carbonates, such as Li$_{0.85}$Na$_{0.61}$K$_{0.54}$CO$_3$, has melting point below 400°C, and provides an opportunity to explore STEP Iron at lower temperatures. At these lower temperature conditions, the electrolysis potential would be expected to be considerably higher. As previously shown the reaction of iron oxide to iron and oxygen is endothermic, with (i) an increase in rest potential with decrease in temperature. This will be exacerbated by (ii) a lower solubility of iron oxide in the eutectic at lower temperature and with lower lithium ion content, and (iii) higher overpotential due to the higher electrolyte resistance of a mixed alkali, compared to pure lithium, electrolyte. The general conditions of electrolysis are similar to those in Table S-1. At the lower 500°C temperature, only a lower concentration of iron oxide could be dissolved in the eutectic and the sustainable current at a reasonable electrolysis potential was only, 0.4 A, rather than 1A. Hence, the electrolysis time was increased from 1 to 2.5 hours to provide a constant total current during the experiment. As summarized in Table S-5, even at the lower current, the lower temperature still requires a high (3.5V) average electrolysis potential, and results in a poor coulombic efficiency. As seen in Table S-5, by 750°C, the $Li_{0.85}Na_{0.61}K_{0.54}CO_3$ eutectic could readily accommodate the full 3 molal $Fe^{3+}$ used in the pure lithium carbonate electrolyte at 800C. It is evident in the figure that in the eutectic at 750C, a higher electrolysis potentials was needed to accommodate the same 1 Amp current used in the pure lithium electrolyte. Interestingly, coulombic efficiencies are high in both cases, and as seen in Fig. S-7, the cathode product contains a low fraction of removed salt, and a high fraction of iron.

**Table S-5.** Effect of eutectic ($Li_xNa_yK_zCO_3$) or pure ($Li_2CO_3$) carbonate, and of temperature, on the electrolytic formation iron. The electrodes are separated by 1.0 cm and are detailed in Table S-1.

| Temperature (°C) | **500** | **750** | **800** |
|---|---|---|---|
| Time of electrolysis (h) | *2.5h* | *1h* | *1h* |
| Electrolysis current (A) | 0.4 | 1.0 | 1.0 |
| Average Potential of electrolysis (V) | 3.5 | 1.95 | 1.69 |
| Carbonate Electrolyte | $Li_{0.85}Na_{0.61}K_{0.54}CO_3$ | **$Li_{0.85}Na_{0.61}K_{0.54}CO_3$** | $Li_2CO_3$ |
| Carbonate Electrolyte: $Li_2CO_3$ (g) | 30.0026 | 25.0002 | 24.9926 |
| $Fe^{3+}$ concentration (mol /kg $Li_2CO_3$) | 0.8 m | 3.0 m | 3.0 m |
| $Fe_2O_3$ weight (g) | 1.9156 | 5.9861 | 5.9843 |
| $Li_2O$ concentration (mol /kg $Li_2CO_3$) | 0.8 m | 3.0 m | 3.0 m |
| $Li_2O$ weight (g) | 0.7177 | 2.2416 | 2.241 |
| Electrolyte Total Weight (g) | 32.6468 | 33.2349 | 33.2174 |
| $Fe^0$ mass in product (g) | 0.026 | 0.388 | 0.381 |
| Coulomb effic-100%x$Fe^0$ mass exp/theory | 3.73 | 55.8 | 54.7 |

**Table S-6.** Effect of planar foil cathode surface area on the electrolytic formation iron. The electrolysis time, current, theoretical maximum mass of iron, and anode are as detailed in Table S-1. Inter-electrode separation is 1.0 cm. Cathodes are described in the table.

| Temperature (°C) | 800 | 800 | 800 | 800 |
|---|---|---|---|---|
| Average Potential of electrolysis (V) | 1.66 | 1.68 | 1.69 | 1.43 |
| Electrolyte Total Weight (g): Electrolyte | 33.2161 | 33.2093 | 33.2174 | 33.281 |
| weighed from a mix of 200.0015g $Li_2CO_3$, 47.8888g $Fe_2O_3$ and 17.9286g $Li_2O$, | | | | |
| $Fe^{3+}$ concentration (mol /kg $Li_2CO_3$) as $Fe_2O_3$ | 3.0 m | 3.0 m | 3.0 m | 3.0 m |
| $Li_2O$ concentration (mol /kg $Li_2CO_3$) | 3.0 m | 3.0 m | 3.0 m | 3.0 m |
| Cathode: Steel shim, 2.5cm×2.5cm | 0.8 cm$^2$ | **2.5 cm$^2$** | **6.25 cm$^2$** | **12.5 cm$^2$** |
| $Fe^0$ mass in product (g) | 0.238 | 0.325 | 0.381 | 0.191 |
| Coulombic efficiency (100%x $Fe^0$ weight experiment/theory) | 34.2 | 46.7 | 54.7 | 27.5 |

Table S-6 summarizes interesting, but unsuccessful, attempts to improve the coulombic efficiency of the planar iron foil cathode, by increasing or decreasing its surface area. As seen in Fig. S-8, two smaller surface area electrodes were compared, as well as a double surface area electrode folded in an accordion configuration to accommodate the double width of the electrode. As summarized in Table S-6, while the larger surface area electrode did decrease the average electrolysis potential, the coulombic efficiency was maximum for the simpler planar, solid cathode. As in prior experiments, the anode was a Ni coil, 16 cm length, 2.0 mm diameter, area 10 cm$^2$, and the anode/cathode inter-electrode separation was 1 cm.

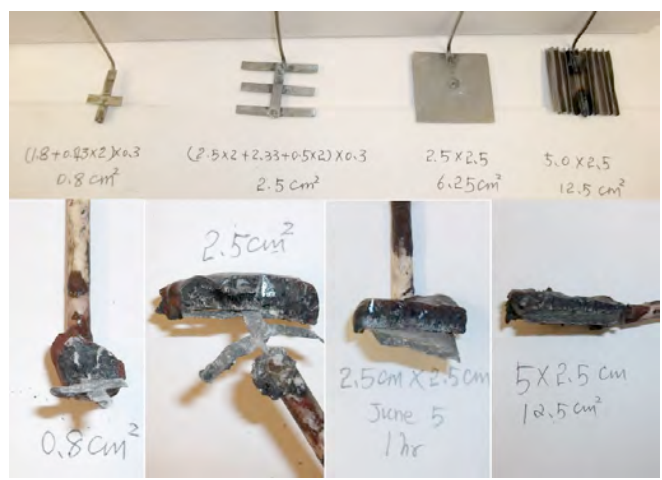

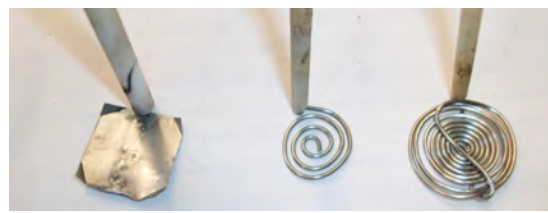

**Fig. S-8 (left)**. Solid steel foil cathodes with various surface areas prior to electrolysis (top) and following electrolysis with partial removal of cathode product (bottom). Spot welds to connect the steel wire contacts are evident.

**Fig. S-9**. The 6.25 cm$^2$ foil, and 5 or 20 cm$^2$ coiled wire, cathodes prior to electrolysis.

Table S-7 summarizes a successful attempt to modify the cathode configuration, which is accomplished by transitioning from a planar, to a coiled, steel cathode. Fig. S-9 compares the planar and coiled cathode configurations. Not shown is the similar, intermediate 10 cm$^2$ coil, which appears to combine the advantages of a loose coil packing with a relatively high surface. Interpolating between the coulombic efficiencies of the 5 or 10 cm$^2$ coiled cathodes, it can be noted that the 6.25 cm$^2$ foil cathode exhibits similar efficiencies to the same surface area coiled wire electrode. A common impurity in iron ores is silicate. Another change in this experiment was addition of 10%, by mass, lithium silicate as an initial attempt to simulate the electrolysis of iron ore with silicate. As seen comparing Tables 6 and 7, the silicate marginally diminishes the coulombic efficiency at the planar, foil electrode to 51%. However, the coulombic efficiency at the 10 cm$^2$ coiled steel wire cathode is higher at 55%.

**Table S-7.** Effect of cathode shape and current density (determined by cathode surface area) on the electrolytic formation of iron. The anode, with an anode/cathode separation of 1.0 cm is detailed in Table S-1. Cathodes are described in the table. Each electrolysis is at 1 A for 2 hours.

| Temperature (°C) | 800 | 800 | 800 | 800 |
|---|---|---|---|---|
| Average Potential of electrolysis (V) | 1.636 | 1.759 | 1.787 | 1.738 |
| Carbonate Electrolyte: Li$_2$CO$_3$ (g) | 25.0 | 25.0 | 25.0 | 25.0 |
| Fe$^{3+}$ concentration (mol /kg Li$_2$CO$_3$) | 3.0 | 3.0 | 3.0 | 3.0 |
| Fe$_2$O$_3$ weight (g) | 5.9885 | 5.9886 | 5.9880 | 5.9883 |
| Li$_4$SiO$_4$ (g) (10 wt % SiO$_2$ content in Fe$_2$O$_3$) | 1.194 | 1.194 | 1.194 | 1.194 |
| Electrolyte Total Weight (g) | 33.9996 | 34.0005 | 34.4254 | 34.0002 |
| Cathode: Area ( cm$^2$) | 6.25 | **5.0** | **10.0** | **20.0** |
| Size: Length×width or diameter ( cm)<br>Shape: foil or coiled wire | 2.5×2.5<br>Fe foil | 13.3×0.12<br>Fe wire | 26.5×0.12<br>Fe wire | 53×0.12<br>Fe wire |
| Current density (mA/cm$^2$) | 160 | **200** | **100** | **50** |
| Fe$^0$ mass in product (g) | 0.7049 | 0.2994 | 0.7693 | 0.6350 |
| Coulombic efficiency (100%x Fe$^0$ weight experiment/theory) | 51% | 22% | 55% | 46% |

In the absence of silicate, the improvement in coulombic efficiency with the larger surface area coiled, rather than smaller surface area planar, cathode is more evident. Table S-8 summarizes results of electrolyses each using a 10 cm$^2$ coiled cathode, and with either 2, 3, 4 or 6 molal Fe$^{3+}$, and Li$_2$O in 800°C molten lithium carbonate. The coulombic efficiency is high and comparable in the 3 and 4 molal electrolytes, with the 3 molal exhibiting a modestly higher efficiency of 70%. In the presence of 10% silicate, as seen in Table S-9, and continuing with use of the preferred coiled cathode, the coulombic efficiency is somewhat higher in the 3, rather than 2, molal Fe$^{3+}$, and higher when a 1:1 equivalent ratio, rather than a 2:1 equivalent ratio of Li$_2$O is used. In all cases the presence of the silicate decreases the measured coulombic efficiency.

**Table S-8.** Effect of Fe$_2$O$_3$ concentration on the electrolytic formation iron. The anode is as detailed in Table S-2, with an anode/cathode separation of 1.0 cm. Cathodes are described in the table. Each electrolysis is at 1 A for 1 hour.

| Temperature (°C) | 800 | 800 | 800 | 800 |
|---|---|---|---|---|
| Average Potential of electrolysis (V) | 1.6645 | 1.826 | 1.847 | 1.584 |

| Carbonate Electrolyte: $Li_2CO_3$ (g) | 26.9996 | 25.0006 | 23.0006 | 21.0003 |
|---|---|---|---|---|
| $Fe^{3+}$ concentration (mol /kg $Li_2CO_3$) | 2.0 | **3.0** | **4.0** | **6.0** |
| $Fe_2O_3$ weight (g) | 4.3117 | 5.9881 | 7.3460 | 10.0603 |
| $Li_2O$ concentration (mol /kg $Li_2CO_3$) | 2.0 | 3.0 | 4.0 | 6.0 |
| $Li_2O$ weight (g) | 1.6134 | 2.2413 | 2.7485 | 3.7643 |
| Electrolyte Total Weight (g) | 32.9352 | 33.2002 | 33.0226 | 34.8159 |
| Cathode: Fe coil, 26.5 cm length, 1.2 mm diameter, area | 10 cm² | 10 cm² | 10 cm² | 10 cm² |
| $Fe^0$ mass in product (g) | 0.7229 | 0.9710 | 0.9496 | 0.4111 |
| Coulombic efficiency (100%x $Fe^0$ weight experiment/theory) | 52% | 70% | 68% | 30% |

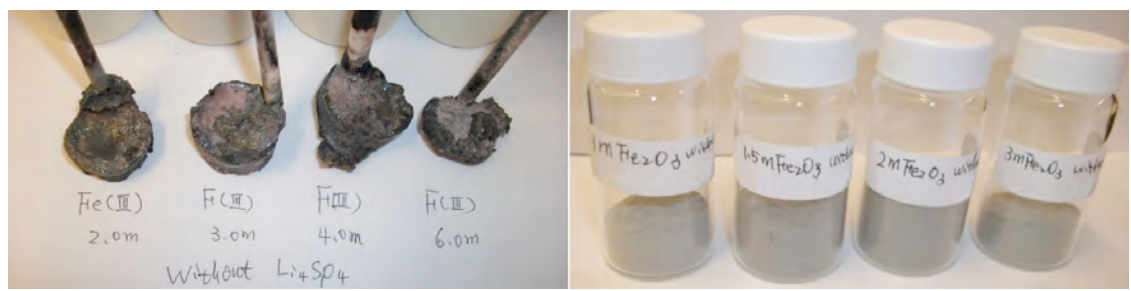

**Fig. S-10**. Left: 10 cm² coiled steel foil cathodes removed after the electrolyses described in Table S-8. Right: removed, then ground product, prior to washing from the electrodes picture on the left side.

**Table S-9.** Effect of $Fe_2O_3$ concentration in an electrolyte containing dissolved silicate (10 wt % $SiO_2$ content in $Fe_2O_3$), on the electrolytic formation of iron. The anode is as detailed in Table S-1, with an anode/cathode separation of 1.0 cm. Cathodes are described in the table. Each electrolysis is at 1 A for 2 hours.

| Temperature (°C) | 800 | 800 | 800 | 800 |
|---|---|---|---|---|
| Average Potential of electrolysis (V) | 1.639 | 1.791 | 1.791 | 1.745 |
| Carbonate Electrolyte: $Li_2CO_3$ (g) | 26.9992 | 24.9994 | 22.9998 | 20.9994 |
| $Fe^{3+}$ concentration (mol /kg $Li_2CO_3$) | **2** | **3** | **2** | **3** |
| $Fe_2O_3$ mass (g) | 4.3115 | 5.9886 | 7.3468 | 10.0603 |

| Li$_2$O concentration (mol /kg Li$_2$CO$_3$) | 2 | 3 | 4 | 6 |
|---|---|---|---|---|
| Li$_2$O weight (g) | 1.6132 | 2.2411 | 2.7485 | 3.7648 |
| Li$_4$SiO$_4$ (g) (10 wt % SiO$_2$ content in Fe$_2$O$_3$) | **0.8605** | **1.1942** | **1.4659** | **2.0068** |
| Electrolyte Total Weight (g) | 33.7870 | 34.4254 | 34.5667 | 36.8151 |
| Cathode: Fe coil, 26.5 cm length, 1.2 mm diameter, area | 10 cm$^2$ | 10 cm$^2$ | 10 cm$^2$ | 10 cm$^2$ |
| Fe$^0$ mass in product (g) | 0.6708 | 0.7693 | 0.4784 | 0.5899 |
| Coulombic efficiency (100%x Fe$^0$ weight experiment/theory) | 48% | 55% | 34% | 42% |

Table S-10 summarizes probes of the temperature effect on STEP Iron electrolysis in a pure lithium carbonate electrolyte (but containing silicate, and 3 molal Fe$^{3+}$ and Li$_2$O). While the electrolysis potential decreases with increasing temperature, the electrolysis efficiency is lowest (23%) at the highest, 900°C, temperature, and highest (58%) at the lowest, 750°C, electrolysis temperature. The lower efficiencies, at higher temperature, may be associated with the greater reactivity and diffusivity of the oxygen produced at the anode, which can back react with iron to form a parasitic iron oxide loss. Table S-11 further probes of silicate effect on STEP Iron. We had previously studied the dissolution of silica, SiO$_2$, as Li$_4$SiO$_4$ in molten carbonates (*10*). Here, we see the trend that higher SiO$_2$ ranging from 10 to 30% (added as a percentage of the iron oxide mass to simulate an impurity in the iron ore), tends to decrease the average electrolysis potential, but also decreases the coulombic efficiency of iron production.

**Table S-10.** Effect of temperature on the 1 A, 2 hour electrolytic formation iron (with silicate in electrolyte). The anode, detailed in Table S-2, is separated from the cathode by 1.0 cm.

| Temperature (°C) | **750** | **800** | **900** |
|---|---|---|---|
| Average Potential of electrolysis (V) | 1.801 | 1.788 | 1.522 |
| Carbonate Electrolyte: Li$_2$CO$_3$ (g) | 25 | 25 | 25 |
| **Fe$^{3+}$ concentration (mol /kg Li$_2$CO$_3$)** | 3 | 3 | 3 |
| Li$_2$O concentration (mol /kg Li$_2$CO$_3$) | 3 | 3 | 3 |
| Li$_4$SiO$_4$ (g | 1.194 | 1.194 | 1.194 |
| Electrolyte Total Weight (g) | 33.9999 | 34.4254 | 33.9998 |
| Cathode: Fe coil, 26.5 cm length, 1.2 mm diameter. area | 10 cm$^2$ | 10 cm$^2$ | 10 cm$^2$ |
| Fe$^0$ mass in product (g) | 0.8067 | 0.7693 | 0.3183 |

| Coulombic efficiency (100%x Fe$^0$ weight experiment/theory) | 58% | 55% | 23% |
|---|---|---|---|

Table S-11. Effect of SiO$_2$ content on the 1 A, 2 hour electrolytic formation iron. Anode/cathode separation is 1.0 cm; anode as detailed in Table S-1.

| Temperature (°C) | 800 | 800 | 800 |
|---|---|---|---|
| Average Potential of electrolysis (V) | 1.794 | 1.812 | 1.672 |
| Carbonate Electrolyte: Li$_2$CO$_3$ (g) | 24.9994 | 25.0000 | 24.9996 |
| Fe$^{3+}$ concentration (mol /kg Li$_2$CO$_3$) | 3 | 3.0 | 3.0 |
| Fe$_2$O$_3$ weight (g) | 5.9886 | 5.9880 | 5.9883 |
| Li$_2$O concentration (mol /kg Li$_2$CO$_3$) | 3.0 | 3.0 | 3.0 |
| Li$_2$O weight (g) | 2.2411 | 2.241 | 2.2410 |
| Li$_4$SiO$_4$ (g) = Fe$_2$O$_3$ weight×SiO$_2$ content÷60.08 (SiO$_2$)×119.84( Li$_4$SiO$_4$) | 1.1942 | 1.7940 | 3.5883 |
| SiO$_2$ content in Fe$_2$O$_3$ ( wt %) | **10%** | **15%** | **30%** |
| Cathode: Fe coil, 26.5 cm length, 1.2 mm diameter, area | 10 cm$^2$ | 10 cm$^2$ | 10 cm$^2$ |
| Fe$^0$ mass in product (g) | 0.7693 | 0.7282 | 0.5668 |
| Coulombic efficiency (100%x Fe$^0$ weight experiment/theory) | 55% | 52% | 41% |

In the next series of experiments the anode stability was improved by raising the anode, from 3 mm below the electrolyte surface (a configuration used in all prior experiments), up to the surface of electrolyte. Prior to this surface anode configuration, anodes occasionally spontaneously broke during the course of the electrolysis. However the surface anodes appear to be fully stable, that is, there is no case of anode discontinuity occurring with the surface anodes in the next 30 experiments, independent of electrolysis conditions, and the anode always appeared to be unaffected by the electrolysis (no corrosion was evident).

Table S-12. Effect of surface area of the anode, when situated at the surface (the interface between the molten electrolyte and the gas above the melt), on the 1 A, 2 hour electrolytic formation iron. Anode and cathode are detailed below in the table.

| Temperature (°C) | 800 | 800 | 800 | 800 | 800 |
|---|---|---|---|---|---|
| Anode: Ni wire, 2.0 mm diameter, length: | 2 cm | 8 cm | 16 cm | 16 cm | 30 cm |
| surface area: | 1.3 cm$^2$ | **5 cm$^2$** | **10 cm$^2$** | **10 cm$^2$** | **18.8 cm$^2$** |

| coil configuration: | tight | tight | tight | loose | tight |
|---|---|---|---|---|---|
| Cathode: Fe coil, l = 26.5 cm, d =1.2 mm, area: | 10 cm$^2$ | 10 cm$^2$ | 10 cm$^2$ | 10 cm$^2$ | 10 cm$^2$ |
| Average Potential of electrolysis (V) | 1.96 | 1.91 | 1.89 | 1.83 | 1.66 |
| Fe$^{3+}$ concentration (mol /kg Li$_2$CO$_3$) as Fe$_2$O$_3$ | 3.0 m | 3.0 m | 3.0 m | 3.0 m | 3.0 m |
| Li$_2$O concentration (mol /kg Li$_2$CO$_3$) | 3.0 m | 3.0 m | 3.0 m | 3.0 m | 3.0 m |
| Electrolyte total mass (g) | 33.2176 | 33.2167 | 33.2172 | 33.2172 | 33.2161 |
| Li$_4$SiO$_4$ (g) | 0 | 0 | 0 | 0 | 0 |
| Fe$^0$ mass in product (g) | 0.490 | 0.826 | 0.808 | 0.563 | 0.655 |
| Coulombic efficiency (100%x Fe$^0$ mass exp/theory) | 35.2 | 59.3 | 58.0 | 40.4 | 47.0 |

In Table S-12 two anode parameters are varied, the surface area, and whether the anode wire is loosely or tightly coiled, as illustrated in the photographs of Fig. S-11. As seen in Table S-12, the electrolysis potential decreases with increasing anode surface area, and its overpotential decreases by 300 mV as the surface area is increase from 1.3 to 19 cm$^2$. With the exception of the largest anode surface cell (which may be an outlier and exhibited an unstable electrolysis potential), the coulombic efficiency increases with increasing anode surface area, and the tightly coiled anode configuration leads to higher efficiency than the loosely coiled anode configuration. Finally, the surface anode in Table S-12, while more stable operates at lower coulombic efficiency than its 10 cm$^2$ tightly coiled counterpart in Table S-9.

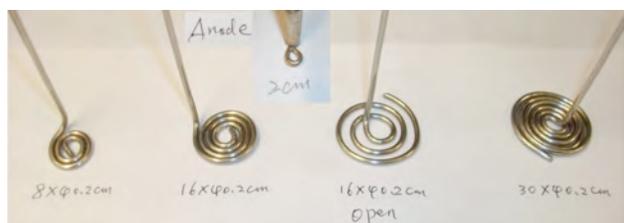

**Fig. S-11**. Coiled nickel wire anodes with a tight or loose (open) prior to the Table S-12 electrolyses.

Table S-13 presents the effect of the gas above the electrolysis on the iron production, when the gas is changed from either air, to carbon dioxide, or to nitrogen. It is seen here that nitrogen significantly lowers the electrolysis potential and that both N$_2$ or CO$_2$ can improve the coulombic efficiency. As previously shown,[1] pure CO$_2$ will be absorbed in the electrolyte according to the back reaction of the lithium carbonate decomposition/reformation equilibrium. Air contains (0.03%) CO$_2$, and molten situated Li$_2$CO$_3$ below a blanket of air will be relatively stable, while pure N$_2$, without CO$_2$, will slowly decompose in accord with the equilibrium equation: Li$_2$CO$_3$ → CO$_2$ +Li$_2$O. Nitrogen above the electrolysis is seen to decrease the electrolysis potential, and has only a marginal impact on coulombic efficiency. As also summarized in Table S-13, additional Li$_2$O added to the molten lithium carbonate electrolyte (above 3 m) decreases coulombic efficiency, but sustains the electrolysis at a lower potential.

**Table S-13**. Effect of the gas above the melt, and the Li$_2$O concentration, on the electrolytic 1A, 2 hour formation of iron.

| Temperature ($^o$C) | 800 | 800 | 800 | 800 | 800 |
|---|---|---|---|---|---|
| Gas above electrolyte | **air** | **air** | **CO$_2$** | **N$_2$** | **air** |
| Anode: Ni wire, 2.0 mm diameter, length: | 16 cm | 16 cm | 16 cm | 16 cm | 16 cm |
| surface area: | 10 cm$^2$ | 10 cm$^2$ | 10 cm$^2$ | 10 cm$^2$ | 10 cm$^2$ |
| coil configuration: | tight | tight | tight | tight | tight |
| Cathode: Fe coil, l = 26.5 cm, d =1.2 mm, area: | 10 cm$^2$ | 10 cm$^2$ | 10 cm$^2$ | 10 cm$^2$ | 10 cm$^2$ |
| Average Potential of electrolysis (V) | 1.87 | 1.80 | 1.95 | 1.62 | 1.71 |
| Fe$^{3+}$ concentration (mol /kg Li$_2$CO$_3$) as Fe$_2$O$_3$ | 3.0 m | 3.0 m | 3.0 m | 3.0 m | 3.0 m |
| Li$_2$O concentration (mol /kg Li$_2$CO$_3$) | **2.0 m** | **3.0 m** | 3.0 m | 3.0 m | **4.0 m** |
| Electrolyte total mass (g) | 32.5140 | 33.2172 | 33.2153 | 33.2131 | 33.9994 |
| Fe$^0$ mass in product (g) | 0.880 | 0.808 | 0.871 | 0.833 | 0.708 |
| Coulombic efficiency (100%x Fe$^0$ mass exp/theory) | 63.17 | 58.0 | 62.5 | 62.8 | 50.8 |

Lowering the electrolysis temperature and decreasing the cathode current density can improve coulombic efficiency. As seen in Table S-14, the columbic efficiency is increased by over 20%, that is to ~85%, by simultaneously decreasing the electrolysis temperature from 800°C to 750°C, and/or by increasing the cathode surface area. This is not observed when the temperature is held constant and the cathode surface area is decreased to 7.5 cm$^2$. Table S-14 onward, are presented in an abbreviated format, without the electrolyte concentration to save manuscript space. In each case the electrolyte total mass is ~ 33 g. In each case the electrolyte is 3.0 m in Fe$^{3+}$ and Li$_2$O and without silicates. The coulombic efficiency also depends on when the electrode is removed from the electrolysis chamber (electrolysis time, Table S-15) and the electrolysis current (Table S-16). As seen in Table S-15, removing the electrode after, 1 hour negatively impacts the efficiency, although this effect presumably may be mitigated if iron oxide is fed into the electrolysis chamber as iron is produced.

**Table S-14**. Effect of the decrease in temperature and variation in cathode surface area on the 1A, 2 hour electrolytic formation of iron with coiled iron cathodes. The electrolyte is 3.0 m in Fe$^{3+}$ and Li$_2$O.

| Temperature ($^o$C) | 750 | **750** | 750 | **800** |
|---|---|---|---|---|
| Gas above electrolyte | N$_2$ | N$_2$ | N$_2$ | N$_2$ |
| Anode: Ni wire, d = 2.0 mm, l = 16 cm, coil, area: | 10 cm$^2$ | 10 cm$^2$ | 10 cm$^2$ | 10 cm$^2$ |
| Cathode: Fe coil, 1.2 mm diameter, length: | 20 cm | 26.5 | 33 cm | 26.5 |

| surface area: | 20 cm² | 10 cm² | 33 cm² | 10 cm² |
|---|---|---|---|---|
| | 7.5 cm² | **10 cm²** | **12.5 cm²** | **10 cm²** |
| Average Potential of electrolysis (V) | 1.90 | 2.1 | 2.0 | 1.62 |
| Coulombic efficiency (100%x Fe⁰ mass exp/theory) | 75.4 | 86.7 | 84.7 | 62.8 |

**Table S-15**. Effect of the electrolysis time in a lower temperature (750 °C) lithium carbonate electrolyte on the electrolytic formation iron with coiled iron cathodes. The electrolyte is 3.0 m in $Fe^{3+}$ and $Li_2O$.

| Temperature (°C) | 750 | 750 | 750 | 750 |
|---|---|---|---|---|
| Gas above electrolyte | $N_2$ | $N_2$ | $N_2$ | $N_2$ |
| Anode: Ni wire, d = 2.0 mm, l = 16 cm, coil, area: | 10 cm² | 10 cm² | 10 cm² | 10 cm² |
| Cathode: Fe coil, l = 26.5 cm, d =1.2 mm, area: | 10 cm² | 10 cm² | 10 cm² | 10 cm² |
| Time of electrolysis (h) at 1.0 amp | *1h* | *2h* | *3h* | *4h* |
| Coulombic efficiency (100%x Fe⁰ mass exp/theory) | 79.6 | 86.7 | 69.6 | 57.4 |

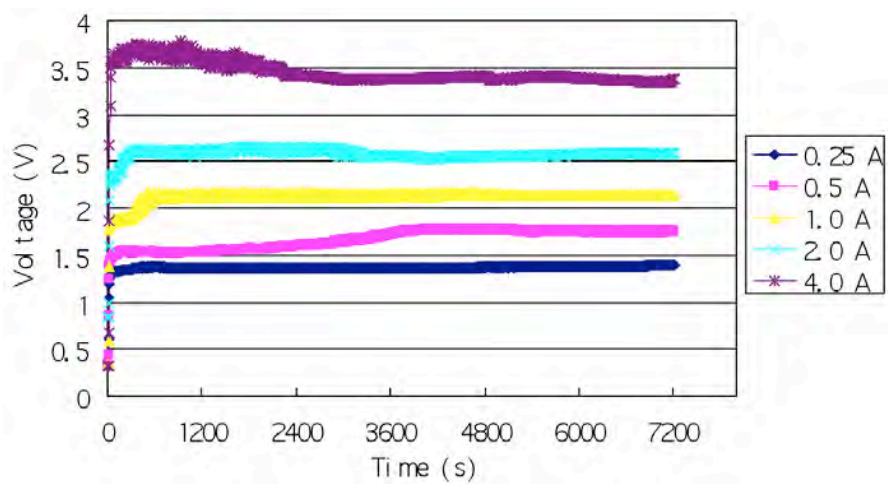

**Fig. S-12**. Variation of STEP Iron electrolysis potential with electrolysis current in a lower temperature (750 °C) lithium carbonate electrolyte.

Current density, Table S-16, substantially effects the electrolysis with a maximum coulombic efficiency of 93% observed at 0.5 A, and as seen in Fig. S-12, a substantial decrease of the electrolysis potential at lower currents; a lowering of 2 volts between the electrolyses at 4.0 A compared to 0.25 A. A further increase in coulombic efficiency of the electrolytic formation of iron at 1 A occurs at even lower temperature, 730°C in Table S-17. This temperature approaches the 723°C melting point of pure $Li_2CO_3$. While the efficiency falls rapidly at high temperature, the electrolysis potential is lower as seen in Fig. S-13. Electrolyte decomposition (from lithium carbonate to lithium oxide and carbon dioxide) occurs more rapidly at higher temperature. There is little decomposition at 750°C,[2] and the decomposition which occured at higher temperature may be controlled or eliminated by increasing the lithium oxide concentration within the electrolyte, or increasing the concentration of carbon dioxide in the atmosphere above the electrolysis. Hence, flowing air (0.03% $CO_2$), rather than nitrogen (0% $CO_2$), above the electrolysis will decrease the rate of electrolyte loss at higher temperature, even though as seen in Table S-17, this nitrogen may marginally improve the coulombic efficiency compared to the electrolysis in air.

**Table S-16**. Effect of the electrolysis current in a lower temperature (750 °C) lithium carbonate electrolyte on the electrolytic formation iron with coiled iron cathodes. Electrolyte is 3.0 m in $Fe^{3+}$ and $Li_2O$.

| Temperature (°C) | 750 | 750 | 750 | 750 | 750 |
|---|---|---|---|---|---|
| Gas above electrolyte | $N_2$ | $N_2$ | $N_2$ | $N_2$ | $N_2$ |
| Anode: Ni wire, d = 2.0 mm, l = 16 cm, coil, area: | 10 cm² | 10 cm² | 10 cm² | 10 cm² | 10 cm² |
| Cathode: Fe coil, l = 26.5 cm, d =1.2 mm, area: | 10 cm² | 10 cm² | 10 cm² | 10 cm² | 10 cm² |
| Electrolysis current (A) during a 2 hour electrolysis | **0.25** | **0.5** | **1.0** | **2.0** | **4.0** |
| Coulombic efficiency (100%x $Fe^0$ mass exp/theory) | 53.7 | 93.1 | 86.7 | 50.3 | 25.6 |

**Table S-17**. Effect of the electrolysis temperature in a lithium carbonate electrolyte on the electrolytic formation iron with coiled <u>horizontal</u> nickel anodes and coiled wire iron cathodes. 1A, 2 hour electrolyses. Electrolyte is 3.0 m in $Fe^{3+}$ and $Li_2O$.

| Temperature (°C) | **730** | **750** | **750** | **800** | **850** |
|---|---|---|---|---|---|
| Gas above electrolyte | $N_2$ | $N_2$ | air | air | air |
| Anode: Ni wire, d = 2.0 mm, l = 16 cm, coil, area: | 10 cm² | 10 cm² | 10 cm² | 10 cm² | 10 cm² |
| Cathode: Fe coil, l = 26.5 cm, d =1.2 mm, area: | 10 cm² | 10 cm² | 10 cm² | 10 cm² | 10 cm² |
| Coulombic efficiency (100%x $Fe^0$ mass exp/theory) | 89.9 | 86.7 | 84.7 | 58.0 | 34.5 |

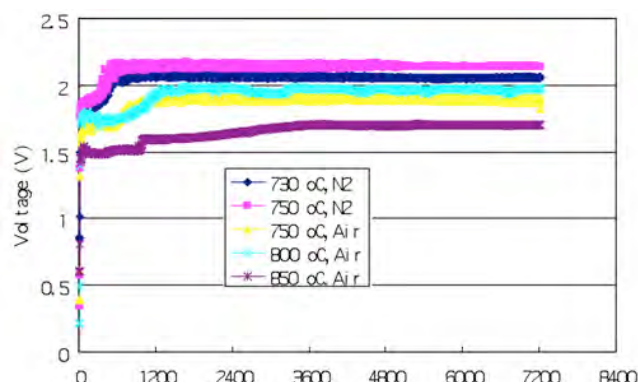

**Fig. S-13**. Variation of STEP Iron electrolysis potential at 1.0 A with electrolysis temperature lithium carbonate electrolyte. Electrolyses are conducted either under air or under nitrogen, as indicated on the figure.

e 32

Table S-18 includes even higher coulombic efficiency STEP iron configurations. The first column contains the same anode on the surface separated by 1 cm from a cathode near the bottom of the cell. As seen in the second column, the use of a smaller (half) diameter nickel or iron wire for the anode and cathode decreases, rather than increases, the coulombic efficiency. We had been working under the hypothesis that hot oxygen would be deleterious to the metallic iron product. Hence, we had previously configured the anode above the cathode to allow gas to evolve without contacting the iron. Interestingly, in Table S-18, at the lower temperature of 730°C in molten lithium carbonate, an opposite, inverted electrode configurations is not only functional, but can exhibit both improved coulombic efficiency and lower electrolysis potential. The coulombic efficiency of these inverted cells is at least 94 to 95% (and this may be considered a lower limit if any iron metal drops into the electrolyte during the cathode removal).

Photos of a vertical cathode inside the anode configuration are presented in Fig. S-14. The potential during electrolysis of these inverted electrode configurations is presented in Fig. S-15. The random oscillations during the cathode on top configuration may be related to a temporary partial blockage of the cathode as anode gas evolved below, passes through this upper electrode. Electrolyte is 3.0 m in $Fe^{3+}$ and $Li_2O$.

**Table S-18**. Effect of the cathode location in molten 730°C lithium carbonate on the electrolytic formation iron. Electrolyte is 3.0 m in $Fe^{3+}$ and $Li_2O$.

| Temperature (°C) | **730** | **730** | **730** | **730** |
|---|---|---|---|---|
| Gas above electrolyte | $N_2$ | $N_2$ | $N_2$ | air |
| Time of electrolysis (h) at 1.0 amp | *2h* | *2h* | *1h* | *1h* |
| Cathode, tight coiled steel wire: Area (cm$^2$) | 10 cm$^2$ | 10 cm$^2$ | 10 cm$^2$ | 10 cm$^2$ |
| Size: Length× or diameter (cm) | 26.5x0.12 | **53x0.06** | 26.5x0.12 | 26.5x0.12 |
| Coil shape: horizontal (plate) or vertical (cylinder) | plate | plate | plate | vertical |
| Cathode above, below, or inside the anode | below | below | **above** | **inside** |
| Anode: Ni coiled wire: Area (cm$^2$) | 10 cm$^2$ | 10 cm$^2$ | 10 cm$^2$ | 10 cm$^2$ |
| Size: wire length × diameter (cm), prior to coiling | 16x0.20 | 32x0.10 | 16x0.20 | 16x0.20 |
| Coil shape: horizontal (plate) or vertical (cylinder) | plate | plate | plate | cylinder |
| Anode above, below, or outside the cathode | above | above | below | outside |
| Average Potential of electrolysis (V) | 1.87 | 1.80 | 1.62 | 1.95 |
| Coulombic efficiency (100%x Fe$^0$ mass exp/theory) | 89.9 | 71.8 | 94.4 | 94.8 |

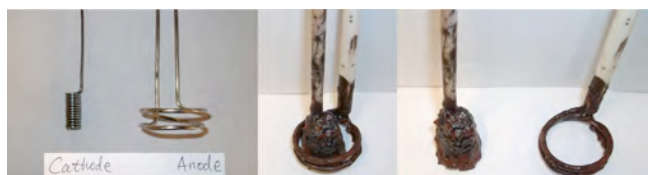

**Fig. S-14**. Vertical cylindrical configured electrodes used in the molten 730°C lithium carbonate electrolysis in which the cathode was placed inside the anode. Left, middle and right-hand photos respectively show the electrodes prior to the electrolysis, after electrolysis and after removal from the electrolyte, and finally after separation of the cathode from the anode.

The ongoing series of STEP Iron electrode electrolyses are each at 730°C, for 2 hours, but are conducted at 0.5, rather than 1.0 A, to probe a path to lower electrolysis potentials, while preserving, or further increasing, the high coulombic efficiencies of iron production. Photographs of these electrodes with various surface areas of the inner, vertical cathode and outer anode coiled electrodes are shown in Fig. S-17. As seen compared to Fig. S-15, in Fig. S-16, the lower current substantially decreases the electrolysis potential, and as seen in Table S-19, retains over 90% the coulombic efficiency. For the same 0.5 A current, a substantially larger surface area electrodes (20 cm$^2$ cathodes and 40 cm$^2$ anodes, providing lower current density conditions), lowers the electrolysis potential to less than 1.4 V as seen in Fig. S-16, but also decreases the coulombic efficiency in the last column of Table S-19.

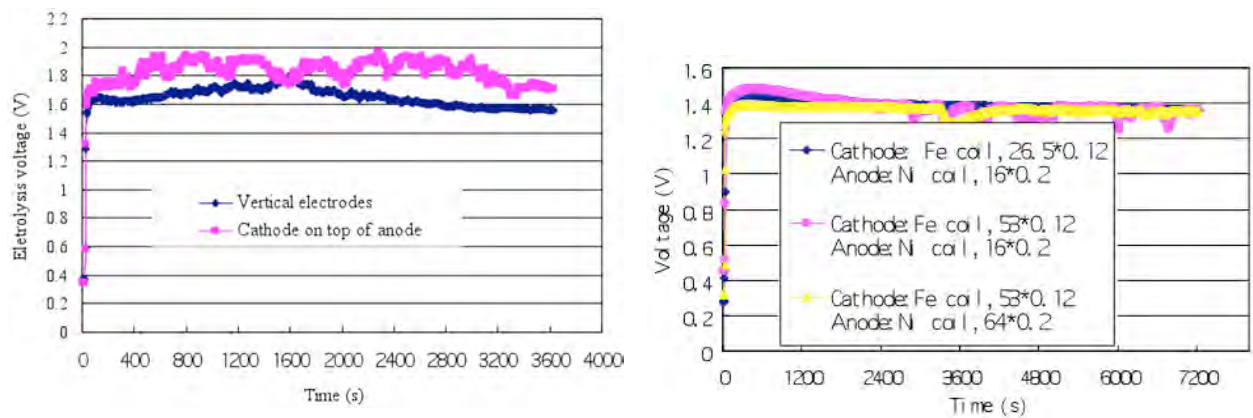

**Fig. S-15 (left)**. Variation of STEP Iron electrolysis potential in a 730°C lithium carbonate electrolyte using the alternate vertical or inverted anode/cathode configurations.

**Fig. S-16 (right)**. Variation of STEP Iron electrolysis potential in a 730°C lithium carbonate electrolyte using the vertical, inner coiled, cathode configuration with different electrode surface areas.

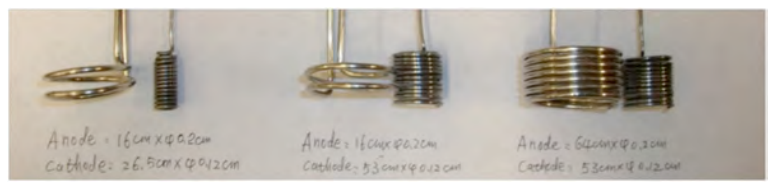

**Fig. S-17**. Vertical cylindrical configured, coiled electrodes with surface area varied by changing the length and coil diameter of the Ni

(outer anode) or steel (inner cathode) wire as used in the molten 730°C lithium carbonate electrolysis.

In an attempt to further prevent any parasitic reaction of the anode and cathode products, the next configuration, places an alumina tube between the outer (anode) and inner vertical, coiled electrodes. This configuration is photographed in Fig. S-18, and although the average electrolysis potential is high at 1.98 V, as seen in the second column of Table S-20, the efficiency does improve. A return to the horizontal, coiled anode (above the cathode and with an alumina separator) retains very high coulombic efficiencies but creates high electrolysis potential. Compared to an average of 1.4 V electrolysis in the Fig. S-17 series of experiments, the next two electrolyses occur at an average, higher electrolysis potentials of 1.81V and 1.78 V respectively. In each case a 10 $cm^2$ cathode has a horizontal configuration and is located below the coiled, horizontal anode. The first utilizes the coiled steel wire and the second a coiled shim (4.8 x 0.65 cm foil) steel cathode. Reflecting the high columbic efficiencies summarized in the last two columns of Table S-18, the high iron content of the product is evident, in Fig. S-19 both at the (uncoiled) shim cathode and still coiled wire cathode. An alternative, horizontal reticulated) Ni anode (consisting of a square 6.25 $cm^2$ area, 0.33 cm thick Ni sponge) above a vertical cathode supported electrolysis at an intermediate potential of 1.51 V. An expanded study of this latter anode, with high surface area morphology, will be presented at a future date.

**Table S-19**. The vertical anode situated outside the cathode configuration, the effect of current and electrode area on iron formation in molten 730°C $Li_2CO_3$ containing 3.0 m $Fe^{3+}$ and $Li_2O$.

| Temperature (°C) | **730** | **730** | **730** |
|---|---|---|---|
| Gas above electrolyte | **air** | **N$_2$** | **N$_2$** |
| Current & time of electrolysis | 1.0A, *1h* | **0.5A, *2h*** | 0.5A, *2h* |
| Cathode, tight coiled steel wire: Area ($cm^2$) | 10 $cm^2$ | 10 $cm^2$ | **20 $cm^2$** |
| Size: Length× or diameter ( cm) | 26.5x0.12 | 26.5x0.12 | **53.5x0.12** |
| Coil shape: vertical (cylinder), inside the anode | | | |
| Anode: Ni coiled wire: Area ($cm^2$) | 10 $cm^2$ | 10 $cm^2$ | **40 $cm^2$** |
| Size: wire length × diameter (cm), prior to coiling | 16x0.20 | 16x0.20 | 64x0.20 |
| Coil shape: vertical (cylinder), outside the cathode | | | |
| Coulombic efficiency (100%x $Fe^0$ mass exp/theory) | 94.8 | 91.2 | 72.8 |

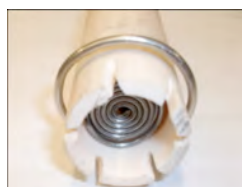
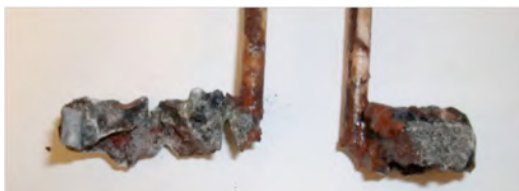

**Fig. S-18 (left)**. Configuration: outer Ni anode is alumina tube shielded & inside is the coiled steel wire cathode.

**Fig. S-19 (right)**. Photo of the high iron content at the (uncoiled) shim and still coiled wire cathodes after removal from the electrolytes summarized in the last two columns of Table S-19 in the molten 730°C lithium carbonate, electrolytic production of iron.

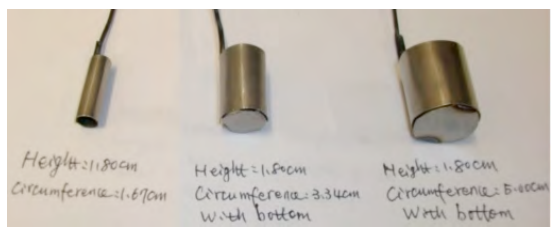

**Fig. S-20**. Vertical steel shim electrodes placed inside the anode for STEP Iron electrolysis as described in Table S-21.

Further increases in coulombic efficiency and decrease in the energy needed to drive the STEP Iron are achieved with relatively small changes to the cathode. As seen in Table S-21, replacement of the vertical inner coiled wire cathode with a similar shaped, steel shim (foil) cathode increases the coulombic efficiency. Furthermore, an increase in anode surface area (40 cm$^2$, compared to the 10 cm$^2$ in the first column of Table S-20), decreases the electrolysis potential to 1.4 V. Further increases in the cathode diameter of these shim electrodes, with or without a "bottom" on the cathode (as pictured in Fig. S-20), decrease the electrolysis potential to less than 1.4 V (Fig. S-21), but, perhaps as a result of the increasing proximity of the anode, results in a decrease in the coulombic efficiency (Table S-21). The time variation of the electrolysis potential for the high columbic efficiency is presented in Fig. S-21 and compared to similar configurations for the production of iron at 730°C in lithium carbonate containing 1.5 m in Fe$_2$O$_3$ and Li$_2$O.

**Table S-20**. The vertical anode inside anode configuration: effect of area and planar versus wire cathodes on the electrolytic formation of iron in molten 730°C Li$_2$CO$_3$ with 3.0 m Fe$^{3+}$ and Li$_2$O.

| Temperature (°C) | **730** | **730** | **730** | **730** |
|---|---|---|---|---|
| Gas above electrolyte | N$_2$ | N$_2$ | N$_2$ | N$_2$ |
| Current & time of electrolysis ( | 0.5A, *2h* | 0.5A, *2h* | 0.5A, *2h* | 0.5A, *2h* |
| Cathode, tight coiled steel wire: Area (cm$^2$) <br> Wire or shim (foil) | 10 cm$^2$ <br> wire | **7.5 cm$^2$** <br> wire in tube <br> (Fig. S-18) | **6.25 cm$^2$** <br> **shim** | 10 cm$^2$ <br> wire |
| Size: Length× wire diameter or shim height (cm) | 26.5x0.12 | **20x0.12** | 4.8x0.65 | 26.5x0.12 |

Page 36

| Coil shape: horizontal (plate) or vertical (cylinder) | vertical | horizontal | vertical | vertical coil id x h: 1.3x0.65 |
|---|---|---|---|---|
| Wire Cathode above, below, or inside the anode | inside | **below** | below | below |
| Anode: Ni coiled wire: Area (cm$^2$) | 10 cm$^2$ | **5.7 cm$^2$** | **10 cm$^2$** | **10 cm$^2$** |
| Size: Length × diameter (cm) | 16x0.20 | 9x0.20 | 32x0.10 | 32x0.10 |
| Coil shape: horizontal (plate) or vertical (cylinder) | cylinder | coil | horizontal | horizontal |
| Anode above, below, or outside the cathode | outside | outside | above | above |
| Coulombic effic. (100%x Fe$^0$ mass exp/theory) | 91.2 | 96.3 | 98.0 | 98.6 |

**Table S-21**. The vertical anode inside anode configuration: effect of electrode shape and area on the electrolytic formation iron at 730°C in molten $Li_2CO_3$ containing 3.0 m $Fe^{3+}$ and $Li_2O$.

| Temperature (°C) | **730** | **730** | **730** | **730** | **730** |
|---|---|---|---|---|---|
| Gas above electrolyte | $N_2$ | $N_2$ | $N_2$ | $N_2$ | $N_2$ |
| Current & time of electrolysis | 0.5A, *2h* | 0.5A, *2h* | 0.5A, *2h* | 0.5A, *2h* | 0.5A, *2h* |
| cathode (inside), vertical cylinder, area (cm$^2$) | 10 cm$^2$ | **3 cm$^2$** | **6 cm$^2$** | **9 cm$^2$** | **3 cm$^2$** |
| coiled steel wire or shim (foil) | wire | **shim** | **shim** | **shim** | **shim** |
| Length× wire diameter or shim height (cm) | 26.5x0.12 | **1.67x1.80** | **3.34x1.80** | **5x1.80** | **1.67x1.80** |
| open or with solid bottom | open | open | bottom | bottom | bottom |
| anode: Ni vertical wire coil or Ni crucible | 10cm$^2$ coil | 40cm$^2$ coil | 40cm$^2$ coil | 40cm$^2$ coil | **crucible** |
| Ni wire prior to coiling: length x diam (cm) | 16x0.20 | 64x0.20 | 64x0.20 | 64x0.20 | |
| coil or crucible diameter: | 4.0 cm | 4.0 cm | 4.0 cm | 4.0 cm | 3.2 cm |
| Coulomb. effic. (100%x Fe$^0$ mass exp/theory) | 91.2 | 100.0 | 97.5 | 47.6 | 98.2 |

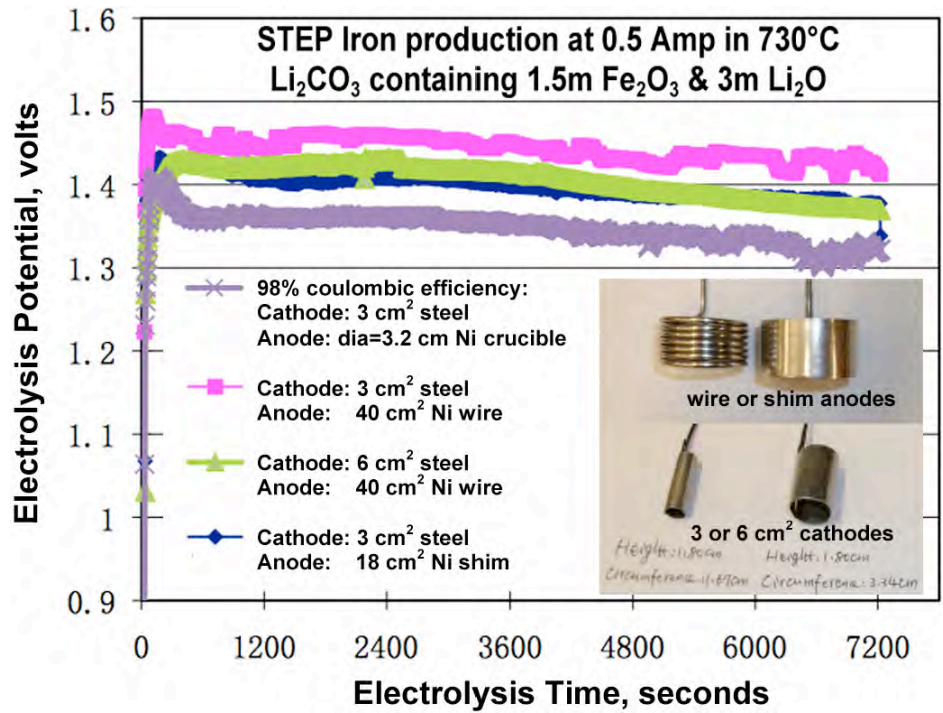

**Fig. S-21.** The variation of the electrolysis potential for the electrolytic production of iron. Inset: photographs of anodes (top) or cathodes (bottom) used in these electrolyses. During the electrolysis the cathode, immersed in the molten electrolyte, is situated within the anode.

The electrolysis configuration is simplified when the electrolysis is conducted in a nickel crucible which comprises both the anode and the cell body in one piece, and can further decrease the electrolysis potential. This configurations of the STEP Iron cell is presented in Fig. S-22; included is a photograph of the cathode after the electrolysis, with the product attached and including some solidified electrolyte). The iron electrolysis product is easy to remove and readily detaches from the cathode. As shown in Fig. S-21, the electrolysis potential is 1.35 V.

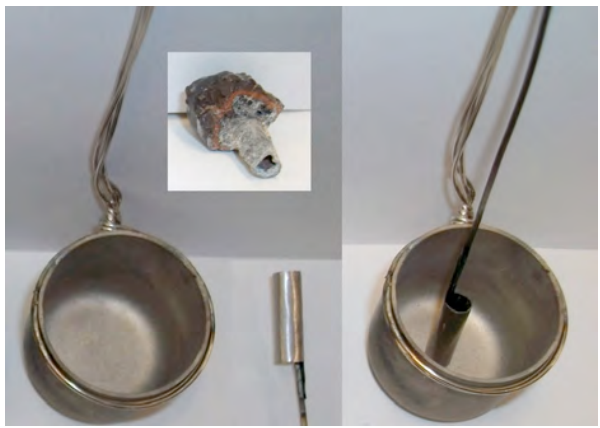

**Fig. S-22.** A nickel crucible serving as both cell wall and anode electrode. A vertical steel cylinder cathode is placed inside the crucible for STEP Iron electrolysis as described in the final column of Table S-21. Inset photo: Cathode with product (and including some solidified electrolyte) after the electrolysis.

This supplementary material provides a detailed characterization of the necessary electrolyte composition, temperature and electrochemical component proximity, size and shape for STEP Iron to provide high coulombic efficiency, at low electrolysis potential, while maintaining kinetically facile electron transfer. The coulombic efficiency approaches 100% yield of the 3 e- reduction of $Fe_2O_3$, The electrolysis potential is less than 1.4 V at a high (current density) rate of iron production. The iron electrolysis product contains iron and electrolyte, and is easy to remove, readily detaching from the cathode. A larger surface area cathode will lower this potential, but may lead to closer proximity to the anode which can decrease the iron coulombic efficiency. Opportunities to significantly decrease the electrolysis potential, while maintaining high cathode current densities, will likely be attained by moving from the largely planar oxygen electrode (anode) to chemically, or mechanically roughened, high microscopic surface area electrodes, and/or using geometric shapes which will expose more of anode surface than the planar electrode.

*Alternative carbonate electrolytes.* A more cost effective solution to the corrosivity of the sodium-potassium STEP carbonate melt (than the use of iridium which is stable as an air electrode during 5 hours of electrolysis in 750°C $Na_{0.23}K_{0.77}CO_3$, compared to nickel air electrodes which corrode) is found by the addition of calcium carbonate or barium salts to the sodium-potassium, lithium-free, carbonate melt (the addition of calcium carbonate is shown here for the electrolytic formation of a carbon product from carbon dioxide splitting, rather than an iron product from iron oxide splittng) (*R-1*). The addition of calcium carbonate can decrease the melting point of a carbonate mix. The sodium/lithium carbonate mix, $Li_{1.07}Na_{0.93}CO_3$, has a melting point of 499°C, but decreases to below 450°C if 2 to 10 mol% equimolar $CaCO_3$ and $BaCO_3$ is added.

In addition to the sodium-potassium carbonate electrolytes, electrolyses are also conducted here in calcium-sodium-potassium electrolytes ranging up to a calcium fraction of $Ca_{0.27}Na_{0.70}K_{0.75}$. Electrodes used are presented in Fig. S-23. A nickel oxygen anode appears to be fully stable during extended (five hour) 0.5 A electrolyses at 750°C in this melt, using the 30 cm$^2$ nickel foil anode and a 7.0 cm$^2$ steel wire cathode, and the electrolysis proceeds at between 1.9 to 2.2V. Unlike the electrolyses conducted in the calcium free (sodium-potassium) carbonate melt, carbon forms and remains on the cathode during electrolysis, and the steel cathode remains the same diameter, as measured subsequent to the electrolysis. As shown subsequent to the electrolyses, in the cathode photographs at the bottom of Fig. S-23, electrolyses conducted in either $Ca_{0.16}Na_{1.03}K_{0.65}CO_3$ or $Ca_{0.27}Na_{0.70}K_{0.75}CO_3$ electrolytes exhibit a thick carbon product on the cathode, while this is not the case following electrolysis without calcium carbonate in $Na_{1.23}K_{0.77}CO_3$. The electrolysis potential and subsequent cathode product, during a repeat of the $Ca_{0.27}Na_{0.70}K_{0.75}CO_3$ electrolysis, but at a constant electrolysis current of 1A, rather than 0.5A, and utilizing a 21cm$^2$ (55 cm coiled steel wire) cathode is presented in Fig. S-24.

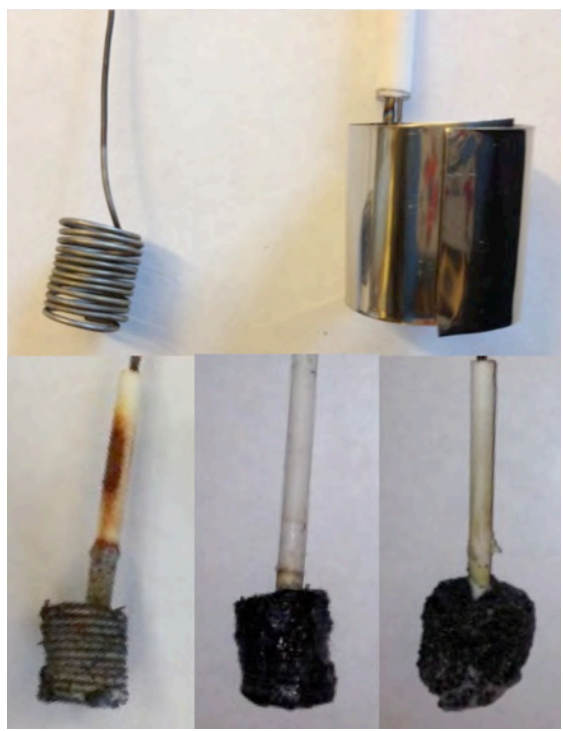

**Fig. S-23.** Top: Cathode (top left) and anode (top right) prior to 0.5 A, 5 hour lithium-free electrolyses at 750°C with increasing calcium carbonate concentration. The cathode is placed inside the anode, which are both immersed in the molten electrolyte. Bottom: Cathodes after electrolysis in lithium-free molten carbonates. Electrolytes used were respectively: $Na_{1.23}K_{0.77}CO_3$ (lower left cathode), $Ca_{0.16}Na_{1.03}K_{0.65}$ (lower middle cathode), and $Ca_{0.27}Na_{0.70}K_{0.75}$ (lower right cathode).

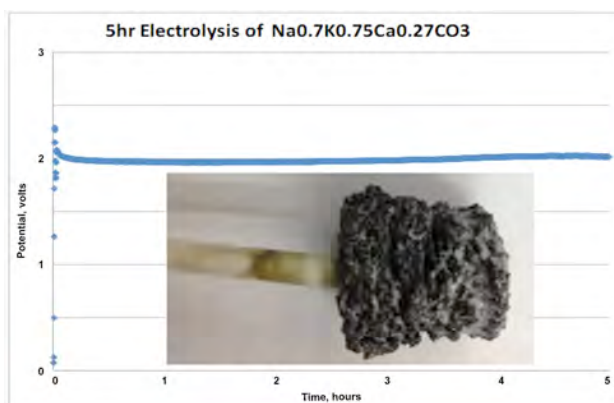

**Fig. S-24.** Time variation of the electrolysis potential during a five hour electrolysis at 1A in $Ca_{0.27}Na_{0.70}K_{0.75}CO_3$ at 750°C. Inset: cathode subsequent to the electrolysis.

**Materials and Methods:** *Chemicals and Materials.* Lithium carbonate utilized is ($Li_2CO_3$, Alfa Aesar, 99%), as well as ferric oxide, $Fe_2O_3$ (99.4%, JT Baker), $Li_2O$ (Alfa 99.5%), 1mm and 2 mm Ni wire (Alfa 99.5 %) and Ni foil (McMaster pure Ni 200 shim), crucibles: nickel (VWR AA35906-KY), Fe wire (Anchor dark annealed annealed), steel foil (McMaster 75 μm 316 steel), crucibles: nickel (VWR AA35906-KY), high purity alumina (AdValue Technology AL-2100), silicon dioxide ($SiO_2$, Spectrum, 325 mesh), lithium orthosilicate is ($Li_4SiO_4$, Alfa Aeasar, 99.9%), Pflatz & Bauer, 99%), boron oxide ($B_2O_3$, 99.98%, Alfa Aesar 89964), vanadium(V) oxide ($V_2O_5$, 99.6%, Alfa Aesar 89964), lithium vanadium oxide ($LiVO_3$, 99.9%, Alfa Aesar 39358) and anhydrous lithium metaborate ($LiBO_2$, 99.9%, Alfa Aesar 12591).

*Electrolyses.* Electrolysis conditions and the systematic variation of the electrolysis cell components are described in the Systematic Optimization of electrolytic iron production in molten carbonate section. The theoretical maximum mass of iron that can be produced from the ferric salt during the electrolysis is calculated as Electrolysis current (A) x Electrolysis time (s) x Atomic Weight Fe / (3 e- x 96,485 As).

*Analyses.* Iron metal is produced by electrolysis in molten carbonate at the cathode. The cathode product is analyzed for iron metal content based on, and improved from the method of Xu and co-workers in which iron metal replaces cupper sulfate, and the product ferrous sulfate is analyzed.[13] The procedure has been further improved by (i) washing the electrolysis product with deionized water, and (ii) replacement of the previous UV/Vis evaluation which was used at the end of the procedure, with a more quantitative (less prone to colorimetric interference) titration by dichromate. The initial rinse removes $Li_2CO_3$ and $Li_2O$ to prevent reaction of $Fe^0$ to form $Fe(OH)_2$ or $Fe(OH)_3$.

In addition to the relative valence state composition of iron, the mass percent of total iron in the sample (including the solidified electrolyte) is shown in the $Fe_{total}$ of column in Table S-1, and the last two columns are measured water soluble and water insoluble mass percent of each sample. The washed, dried insoluble component consists primarily of iron (iron metal and iron oxides). The iron analysis of weighed samples from each layer yields the concentrations: $[Fe_{total}]$, $[Fe^\circ]$, $[Fe^{2+}]$, and $[Fe^{3+}]$ (the latter concentration is determined from the difference of $[Fe_{total}]$-$[Fe^\circ]$-$[Fe^{2+}]$). A separate, weighed sample from each layer is washed, followed by subtraction of the mass of the dried insoluble (filtered, dried) component. This yields the mass of the remaining soluble components. The water soluble salts components consist of $Li_2CO_3$ and $Li_2O$. The $Li_2O$ dissolves as LiOH, including $Li_2O$ stripped from lithiated iron oxides when brought in contact with the wash water (*e.g.* $LiFeO_2 + H_2O$ liberates soluble LiOH), and the 105°C dried components are weighed as the insoluble salts).

*The analysis procedure for iron metal is:*

1) The cathode product is ground in a mortar and pestle, until it can be sifted and dispersed through a 70 mesh (212 μm) sieve.

2) The ground product is washed with deionized water, then extracted by suction filtration and rinsing the precipitate with deionized water until pH is near 7. The precipitates, residue and filter paper are collected to react with $CuSO_4$.

3) To 0.5 g of the ground product is added 50 ml of 0.5 M $CuSO_4$, to form:

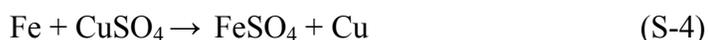
$$Fe + CuSO_4 \rightarrow FeSO_4 + Cu \qquad (S-4)$$

4) After boiling this stirred solution for 1 hour, it is immediately filtered (to prevent the reaction of $O_2$ with $Fe^{2+}$) with a GF/A (Whatman glass microfiber) filter paper into a 250 ml volumetric flask, and the filter paper is washed with double deionized (18 MΩ) water also into the flask, and diluted to 250 ml.

5) 25 ml of the 250 ml filtrate is sampled by pipette into a 250 ml erlenmeyer flask,

and the following solutions are added to the flask: 20 ml of "A", 20 ml of "B", 50 ml of water, and 3 drops of indicator solution "C", where

A: is a mix of 50 ml of water with 10 ml concentrated $H_2SO_4$

B: 700 ml of water with 150 ml concentrated $H_2SO_4$, 150 ml $H_3PO_4$ (binds colored $Fe^{3+}$, which is colored, as colorless $Fe(HPO_4)_2^-$, to improve clarity of the endpoint)

C: Is the indicator solution consisting of 0.2% aqueous Diphenlyamine 4-sulfonic acid sodium salt

D: Is the titrant consisting of 0.004167 M (6 x dilution of 0.025 M) $K_2Cr_2O_7$

which tritrates as 1 equivalent $K_2Cr_2O_7$ per $FeSO_4$; each ml of solution D = 1.3962 mg of $Fe^\circ$ metal.

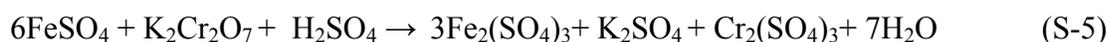
$$6FeSO_4 + K_2Cr_2O_7 + H_2SO_4 \rightarrow 3Fe_2(SO_4)_3 + K_2SO_4 + Cr_2(SO_4)_3 + 7H_2O \qquad (S-5)$$

The endpoint is observed as a color change from light blue (initial) to the endpoint's purple.

This titration analysis is also confirmed by weighing the mass of magnetically removed the iron product that was washed & dried to remove oxide. The reliability of the analysis during the titration is improved when solutions are stirred slowly in step 2, rather than rapidly, to prevent significant errors (underestimating the Fe° content of the product with increasing speed of stirring) due to the introduction of oxygen, which can convert ferrous to ferric prior to the titration. The need to switch to a lower stirring speed for the Fe° analysis was discovered and applied to the latter half of the experiments in this study. Under this latter condition replicate analyses of Fe° metal mass from are reproducible to within a $\pm 2\%$.

*The analysis procedure for total Fe is:*

In accord with the method of Shi et al (*R-2*), a 0.1g sample was placed in a 250 ml flask with the addition of 20 ml of 1:1 diluted HCl. The sample was placed on a mixer hot plate at about 90 °C (to prevent volatile loss of $FeCl_3$ at higher temperature) until completely dissolved. 20 ml water was added under $N_2$ (99.999%). Al powder was added in three 0.1 g portions (in large excess of the theoretical amount to reduce $Fe^{3+}$ to $Fe^{2+}$). With a minimum of stirring, the Al quickly reacts with $Fe^{3+}$ and $H^+$ to form $Al^{3+}$, and $Fe^{3+}$ was reduced to $Fe^{2+}$. A color change from yellow to light yellow was observed until the solution was transparent. The analysis occurs in accord with the following equations:

$$Al + Fe^{3+} \rightarrow Al^{3+} + Fe \qquad (S\text{-}6)$$

$$Fe + 2H^+ \rightarrow Fe^{2+} + H_2 \qquad (S\text{-}7)$$

100 ml of $H_2O$ was added and the solution was cooled until room temperature. 20 ml sulfuric-phosphoric acid solution and 5 drops of the diphenylamine indicator where added after which the solution was titrated with the standard $K_2Cr_2O_7$ solution (0.0250 mol 1/6 $K_2Cr_2O_7$/L).

Total $Fe^{3+}$ (%) = V×N× FW Fe ÷1000/S

V-- standard $K_2Cr_2O_7$ solution volume, ml.

N-- standard $K_2Cr_2O_7$ solution concentration, mol/L.

FW Fe = 55.85 g/ mol.

S—specimen weight, g.

*The analysis procedure for $Fe^{2+}$* is based on the method of reference R-2, as modified by references R-3 and R-4. In this analysis, 0.5 g of the sample to be analyzed was added to 250 ml flask and 20 ml of 1:1 diluted HCl was added. This solution was mixed on a hot plate under 99.999% $N_2$ at 90 °C until completely dissolved. 100 ml $H_2O$ was added and the solution allowed to cool to room temperature. 20 ml sulfuric-phosphoric acid solution (prep: add 600 ml concentrated $H_2SO_4$ to 800 ml stirred DI, then add 600 ml of 86% phosphoric acid) and 5 drops of the diphenylamine indicator were added and titrated with the standard $K_2Cr_2O_7$ solution (0.0250 mol 1/6 $K_2Cr_2O_7$/L) to a sharp endpoint color change from green to purple.

In accord with the equation:

$Fe^{2+}$ (%) = V×N×FW Fe÷1000/S.

V-- standard $K_2Cr_2O_7$ solution volume, ml.

N-- standard $K_2Cr_2O_7$ solution concentration, mol/L.

S—specimen weight, g.

*The analysis procedure for $Fe^{3+}$* is the straightforward difference from the known total iron, the sum of the iron metal and ferrous species as: $Fe^{3+}$(%)= Total Fe(%)-$Fe^0$ (%)-$Fe^{2+}$ (%)

*The analysis procedure for the aqueous soluble components:*

Filters were dried in an oven at 105 °C for 1 hour, and then cooled in a desiccator, and weighed. After weighing, samples (~0.3g) to be analyzed were stirred one hour in 100 ml of DI (18 MΩ) in a 250 ml flask at room temperature, then filtered using suction, and the filtrate washed on the filter with three 20 mL volumes of DI water. The filter is transferred onto a glass weighing dish, and dried 105°C for a minimum of 1 hour until a minimum mass is measured (after cooling to room temperature in the desiccator).

Equation:   Water soluble substances (wt%) = (A-B)×100/A            (S-8)

A—Specimen weight, g.

B-- Residue on the filter paper weight, g.

**References**

R-1 S. Licht, Stabilization of STEP electrolyses in lithium-free molten carbonates. *arXIV,* **arXiv**: 1209.3512 [physics.chem-ph], Sept. 18, 1-4 (2012).

R-2 X. Shi, J. Liao and S. Wang, *Rock and Mineral Analysis* **28**, 197 (2009).

R-3 ASTM designation: D3872 – 86. Standard Test Method for Ferrous Iron in Iron Oxides. *Annual Book of ASTM Standards* 1 (1999).

R-4 Z. Xu, H. Hwant, R. Greenlund, X. Huang, J. Luo and S. Anshuetz, *J Min. Mat. Characterization & Eng.* **2**, 65 (2003).